\documentstyle[11pt,aaspp4]{article}

\newcommand{\beq}{\begin{equation}}
\newcommand{\eeq}{\end{equation}}
\newcommand{\lb}{\langle}
\newcommand{\rb}{\rangle}

\def\ga{\mathrel{\hbox{\rlap{\hbox{\lower4pt\hbox{$\sim$}}}\hbox{$>$}}}}
\def\la{\mathrel{\hbox{\rlap{\hbox{\lower4pt\hbox{$\sim$}}}\hbox{$<$}}}}
\def\ap{\mathrel{\hbox{\rlap{\hbox{\lower4pt\hbox{$\sim$}}}\hbox{$\propto$}}}}
\def\comb{\prod}

\begin{document}

\title{A Numerical Renormalization Solution for \\ Self-Similar
Cosmic Structure Formation}

\author{H. M. P. Couchman}
\affil{Department of Physics and Astronomy, University of Western Ontario, 
	London, ON, N6A 3K7}
\authoremail{couchman@coho.astro.uwo.ca}

\author{P. J. E. Peebles}
\affil{Joseph Henry Laboratories, Princeton University,
	Princeton, NJ, 08544}
\authoremail{pjep@pupgg.princeton.edu}

\begin{abstract}
We present results of a numerical renormalization 
approximation to the self-similar growth of clustering
of a collisionless pressureless fluid out of a power-law spectrum 
of primeval Gaussian mass density fluctuations, 
$P(k)\propto k^n$, in an Einstein-de~Sitter cosmological model.
The self-similar position two-point correlation function,
$\xi (r)$, seems to be well established. The
renormalization solutions for $\xi (r)$ show a satisfying
insensitivity to the parameters in the method, and at 
$n=-1$ and $n=0$ are quite close to the Hamilton {\it et al.}
formula for interpolation between the large-scale perturbative
limit and stable small-scale clustering. The solutions are 
tested by the comparison of the mean relative 
peculiar velocity $\lb v_{ij}\rb$ of particle pairs $(ij)$ 
and the velocity derived from $\xi (r)$ under the assumption of 
self-similar evolution. Both the renormalization 
and a comparison conventional N-body solution are 
in reasonable agreement with the test, although the 
conventional approach does slightly better at large separations
and the renormalization approach slightly better at small
separations. Other comparisons of renormalization and
conventional solutions are more demanding and the results much
less satisfactory. Maps of the particle 
positions in redshift space in the renormalization solutions show
more nearly empty voids and less prominent walls
than do comparison conventional N-body solutions. 
The rms relative velocity dispersion is systematically smaller in
the renormalization solution; 
the difference approaches a factor of two on small scales. There
also
are substantial differences in the frequency distributions of clump
masses in renormalization and conventional solutions. The third 
moment $S_3$ from the distribution of mass within cells is in
reasonable agreement with second-order perturbation
theory on large scales, while on scales less than the clustering 
length $S_3$ is roughly consistent with hierarchical
clustering but is heavily affected by shot noise.

\end{abstract}

\section{Introduction}

Clustering in a self-similar cosmogony is of 
interest because the simple physics might allow particularly
accurate solutions. The reliability of a numerical approximation
to a self-similar solution may be tested by the required 
scaling of properties with time and, in limiting cases, by
comparison to analytic solutions. A reliable numerical
self-similar solution in turn may be important as a benchmark for
more realistic models for cosmic structure and as a guide to the
formulation of analytic 
theories for self-similar evolution, as in the search
for a closure {\it ansatz\/} for the BBGKY hierarchy 
(eg. Davis \&\ Peebles 1977; Ruamsuwan \&\ Fry 1992).

There can be aspects of a self-similar solution
that are more accurately obtained by conventional methods than by 
the renormalization approach used here. An example is the mean
relative proper peculiar velocity on large scales: we find the
conventional solutions better fit the velocity derived from the
two-point correlation function $\xi (r)$. Other discrepancies
between renormalization and conventional solutions may be the
fault of the conventional approach. 
An example found here is the frequency distribution of
cluster masses on scales small compared to the clustering length.
A different application of conventional methods might do better,
of course, but until this can be established and the problem
identified and remedied such discrepancies certainly indicate the
need for caution in the application of numerical simulations.

There have been impressive advances in numerical N-body
computations. For example, an early study of a self-similar 
solution used $N=90$ particles
(Peebles 1971); Peebles \& Groth (1976) reached $N=2000$;
Aarseth, Gott, \& Turner (1979), $N=4000$; Efstathiou \&\
Eastwood (1981), $N=20,000$; and Efstathiou, Frenk, White, \&\
Davis (1988), $N=32^3$. More recent studies of
numerical self-similar solutions use $N=64^3$ (Colombi, Bouchet,
\&\ Hernquist 1996) and $N=128^3$ (Jain, Mo, \&\ White 1995; Yess
\& Shandarin 1996), and
solutions with initial conditions that are not quite
scale-invariant reach $N=256^3$ (Thomas {\it et al.}~1997) and
$N=288^3$ (Jain \&\ Bertschinger 1994). The
requirements of a numerical self-similar solution are 
demanding, however. The clustering length $r_o$ (at which the rms
density contrast is close to unity) has to be much smaller than
the size of the system. In a conventional solution the initial 
conditions are quite unrealistic on the scale of the interparticle 
separation, and this could compromise the clustering that
develops on smaller comoving scales (Splinter et al. 1997). In a
cube of unit width this would mean we want \beq N^{-1/3}\ll r_o\ll 1 .\eeq
Even at $N=256^3$ there is not much room to test that the
clustering pattern has forgotten transients from the
unrealistic initial conditions and has approached self-similar 
evolution. 

A numerical renormalization scheme (Peebles 1985) offers a useful
check on the possible effect of the limited range of the
clustering length $r_o$ in the conventional N-body approach. The
renormalization method addresses the problem by a repeated
rescaling that keeps $r_o$ small compared to the size of the
system. The first application of the method used 
only $N=1000$ particles, and it certainly is timely to reconsider
the approach with modern numerical techniques. This study mainly
uses $N=64^3$ particles, and we compare the renormalization
solution to a conventional computation with the same particle
number. 

In \S 2 we review the definition of the self-similar clustering
problem. 
The great increase in the particle number allowed by present
technology permits a more careful treatment of the large-scale
density fluctuations introduced at each renormalization step. A
new approach is
presented in \S 3 along with a summary of the other steps in the
renormalization method. Section 4 lists parameters for the
computations. Solutions are obtained for two initial conditions,
$n = -1$ and $n=0$, in the primeval mass density fluctuation spectrum 
\beq P(k)\propto k^n.\label{eq:P(k)} \eeq The case $n=-1$ gives a
not unreasonable first approximation to the galaxy distribution,
although the relative velocity dispersion is too large because of
the large mass density in the Einstein-de~Sitter model. The case
$n=0$ offers a useful comparison. The numerical results presented
in \S 5 are limited to a few commonly discussed statistics,
including second-order moments in velocity and elements of
third order in position.

\section{Self-Similar Gravitational Clustering}

Self-similar gravitational clustering in an expanding world
model has no fixed characteristic length: the distribution and
motion of the matter at world times $t_1$ and $t_2$ are
statistically the same after lengths $r_1$ and $r_2$ at the two
epochs $t_1$ and $t_2$ are scaled by the relation 
\beq
   r_2 = r_1(t_2/t_1)^\alpha .\label{eq:alpha}
\eeq
The index $\alpha$ is a constant, and coordinate lengths $r$ 
comove with the background cosmological model. 
The gravitational interaction of the matter is treated in the
nonrelativistic Newtonian limit. This requires that peculiar
velocities are much smaller than the velocity of light (and, which
is almost equivalent, that the magnitude of the Newtonian
potential of the departure of the mass distribution from
homogeneity is much less than $c^2$), but otherwise allows 
arbitrarily strong nonlinear clustering. We use the
Einstein-de~Sitter model with negligible pressure, where the
expansion parameter scales with proper world time as 
\beq
	a\propto t^{2/3}.\label{eq:EdeS}
\eeq
The power spectrum $P(k)$ of the 
mass distribution on scales large compared to the nonlinear
clustering length is a power law (eq.~[\ref{eq:P(k)}]).
If the clustering scale is growing in comoving units then this
power law applies on any comoving length scale at small enough
expansion parameter. In this sense equation~(\ref{eq:P(k)}) is
the primeval mass fluctuation spectrum. The primeval fluctuations 
from homogeneity are assumed to be a random Gaussian process
determined by the parameter $n$ in $P(k)$.

Matter is treated as a pressureless collisionless fluid.
In linear perturbation theory the density contrast of the
mass distribution averaged over comoving scale $r$ varies as 
\beq
	\delta\rho /\rho =\delta ({\bf r},t)\propto
		t^{2/3}r^{-(3+n)/2}. \label{eq:pert}
\eeq
The crossing of orbits produces a coarse-grain
average pressure, but this is the sum of noninteracting
pressureless components with different velocities and
mass densities. The effective Jeans length of the coarse-grain
pressure must be on the order of the clustering length $r_o(t)$
defined by the mass autocorrelation function: 
\beq
	\xi (r) = \lb\rho ({\bf y})\rho ({\bf r}+{\bf y})\rb 
		/\lb\rho\rb ^2 - 1;
	\qquad \xi (r_o) = 1. \label{eq:r_o}
\eeq
This clustering length must scale with all others 
(eq. [\ref{eq:alpha}]):
\beq
   r_o\propto t^\alpha\propto t^{4/(9+3n)}\propto a^{2/(3+n)},
		\label{eq:r_o-scaling}
\eeq
where the second expression follows from equation~(\ref{eq:pert}).
The relation between the power spectrum index~$n$
and the length scaling index $\alpha$ is 
thus (Peebles 1965)
\beq
	\alpha = 4/(9+3n). \label{eq:alpha(n)}
\eeq
If $n>-3$ the Fourier transform of $P(k)$, which is the mass
autocorrelation function, is defined at large separation, and
equation~(\ref{eq:alpha(n)}) indicates $\alpha >0$, meaning the
mass within the scale of nonlinear clustering increases with
increasing time, as wanted. Equation~(\ref{eq:pert}) fails
at $n>4$, where the power spectrum on large scales is dominated
by the coupling to small-scale nonlinear clustering (Peebles
1980, \S28). Thus it is reasonable to seek solutions for the
power law index in the range
\beq
	-3<n<4.
\eeq

Perhaps the most interesting generalization of the cosmological
self-similar clustering problem would be to primeval density
fluctuations that  
are scale-invariant but not Gaussian, that is, fractal.

There is increasing evidence that 
the mass density in matter capable of clustering is less than the
Einstein-de~Sitter value (Peebles 1997; Bahcall 1997). If the
mass density is 
low the self-similar solution still may be useful as a 
description of conditions at large redshift, when the density
parameter was close to unity, and as an initial 
condition for numerical integration to the present.

This discussion assumes that the self-similar solution exists and is
unique, in the sense that Gaussian initial conditions with
given power law index $n$ evolve to a definite
set of $n$-point correlation functions
of the scaled lengths $x=r/t^\alpha$ and streaming velocities
$v/t^{(\alpha - 1/3)}$, and that the moments of the
distributions of mass and momentum within cells are
finite. We know of no
evidence against this assumption but it seems prudent to bear the
issue in mind.

\section{The Numerical Renormalization Method}

We summarize the elements of the numerical renormalization
method, more details of which are in Peebles~(1985), and then
describe our new treatment of the primeval density fluctuations.

\subsection{The Procedure}

The model uses the motion of $N$ particles in a cube with fixed
comoving width and periodic boundary conditions,
and iterates through
the following steps. First, the equation of motion of the particles
under their mutual gravitational interaction is integrated forward in
time until the proper width of the cube has increased by the factor
\beq a_f/a_i=2^{(3+n)/2} = a_{\rm max}.\eeq This increases the
comoving clustering length $r_o$ by a factor of two
(eq.~[\ref{eq:r_o-scaling}]). Second, to bring the ratio of the
clustering length to the cube width back to the original value, adjoin
eight copies of the particle positions and velocities to make a cube
eight times the comoving volume of the original. Third, to scale all
characteristic quantities back to the original values change units of
length, time, and mass by the factors \beq r\rightarrow r/2,\qquad
a\rightarrow a/2^{(3+n)/2},\qquad t\rightarrow t/2^{(9+3n)/4}, \qquad
m\rightarrow m/8.  \eeq 
This scales proper peculiar velocities as
$v\rightarrow v/2^{(1-n)/4}$. 
Fourth, fuse particles to bring the
number back to the original value. This is accomplished by
placing at random one in eight of the particles 
with its original mass and velocity at its place in each of
the adjoined cubes. That is, with 50\%\ probability each
Cartesian position component of each particle is shifted by the
width of the original cube. The Fourier modes on the scale of
the box are depopulated by the renormalization. The final step
is to repopulate these Fourier modes by shifting
particle positions 
and velocities in a manner appropriate for the desired input
spectrum. This step has been improved compared with the 
previous treatment, as discussed in \S 3.2.

The fourth step, the reduction from $8N$
to $N$ particles by the selection of the position and scaled
velocity of one particle, is statistically unbiased 
(Peebles 1985) but crude. It is not easily 
improved, however. It would be straightforward to  
identify neighboring groups of eight
particles, and sensible to replace such a group with a single
particle at the center of mass, but difficult to find 
a reasonable prescription for the velocity of the merged
particle. The appropriate 
velocity would be that of the center of mass if the eight
particles happened to be a gravitationally 
bound group. But if the group were selected from a small section
of a virialized halo the center of mass velocity would be biased
low and the selection of the scaled velocity of one of the
particles in the group would keep the halo virialized. It would
be simple and likely beneficial to replace the particle
velocities in an unusually tight gravitationally bound pair (or
group) with the center of mass velocity, just prior to the fourth
step. This has not been done but might be considered in future
applications of the renormalization method.

\subsection{Application of Large-Scale Density Fluctuations}

The operation described here applies mass density fluctuations on the
scale of the renormalised cube so as to mimic
the assumed primeval Gaussian random process.

We denote each Fourier component, or mode, in the periodic space
defined by the cube by the 
comoving wavenumber ${\bf k} =2\pi{\bf m}$, where {\bf m} is a
triplet of integers which may be positive or negative. Thus the
fundamental mode in the direction of the $x$-axis has
${\bf m}=(1,0,0)$. Let the mode amplitude immediately after a
renormalization step be $\tilde\delta ({\bf k})$. In linear perturbation
theory the integration of the equation of  
motion of the particles from $a_i=1$ to $a_f=a_{\rm max}$
preserves the comoving wavenumber and brings the 
amplitude to $\tilde\delta ({\bf k})a_{\rm max}$. The next
renormalization doubles the wavenumber of this mode, whilst
leaving the amplitude unchanged (apart from the introduction of
Poisson noise). Further iterations map the mode to wavenumbers
$2^j\bf k$ with amplitude $\propto a_{\rm max}^j$. 

We choose the set of modes, ${\bf k}_j$, $j=1, \ldots, l$, to be added
to the cube after each renormalization step so that no new mode
corresponds to one populated by the mapping ${\bf k}\rightarrow
2{\bf k}$ from any previous iteration. To achieve this it is
sufficient to populate any mode for which $\comb({\bf k}/2)=0$,
where $\comb$ is the three-dimensional comb such that 
$\comb({\bf k})=1$ if ${\bf k}/(2\pi )$ is an integer triple and
vanishes otherwise. 
The added modes have amplitude $A({\bf k})\propto k^{n/2}$
and randomly chosen
phases. In linear perturbation theory, after a series of $J$ 
iterations the spectrum of modes has the form
\beq 
	\tilde\delta({\bf k})=
	\sum_{j=0}^{J-1} A({\bf k}/2^j) 
	e^{i\phi_{\bf k}}\, a_{\rm max}^j \comb({\bf k}/2^j),
			\label{eq:c1}
\eeq
where $\phi_{\bf k}$ is a
uniform random variate in $[0,2\pi ]$, and the phases are
constrained such that $\tilde\delta(-{\bf k}) =
\tilde\delta^*({\bf k})$ to ensure a real density displacement.
Equation~(\ref{eq:c1}) corresponds to the sum over a series of
waves with decreasing density in
wavenumber space and correspondingly increasing values of the
amplitudes, $\propto a_{\rm max}^j$ . 

Our choice for the distribution of applied modes is \beq A({\bf
k})=\cases{ \beta\, k^{n/2} \bigl(1-\comb({\bf k}/2)\bigr),& $1\le
\Bigl({\textstyle k\over 2\pi}\Bigr)^2\le \Bigl({\textstyle k_c\over
2\pi}\Bigr)^2 = 22$ \cr \noalign{\smallskip} 0, & otherwise.},
\label{eq:c2} \eeq where $\beta$ is a normalization constant.  This
adds 202 new (independent) modes at each iteration. The
minimum-wavelength applied modes have ${\bf m}=(3,3,2)$ and the allowed
permutations of this triplet. This corresponds to a wavelength of 0.21
of the box width. The choice of $k_c$ is meant to satisfy
two conditions: that we add modes only at relatively large scales,
where the mass density fluctuations are close to linear, and that we
populate a large number of modes so that the input fluctuations are
not dominated by a small number of modes. In the treatment of
Peebles~(1985) modes, ${\bf m}$, were added with
$m_{x,y,z}\in(-1,0,1)$ and ${\bf m}\ne0$. With the requirement that
the Fourier modes be Hermitian this corresponds to 13 new independent
waves added at each renormalization.

The applied density fluctuations are close to Gaussian because they
are the sum of a significant number of independent plane waves. There
are unrealistic gaps in the spectrum, however; this is illustrated in
Figure~\ref{fig0}, which shows the wave numbers and amplitudes
$|\delta _{\rm k}|$ for $n=-1$ applied at each iteration, and evolved
forward in time in linear perturbation theory as described by
equations~(\ref{eq:c1}) and~(\ref{eq:c2}). The scatter plot of the
Fourier amplitudes of the mass distribution after the integration step
of a stable renormalization solution shows traces of the bands in
Figure~\ref{fig0}, along with the dense distribution of points at high
wavenumber resulting from nonlinear evolution. The spread in
Fourier mode amplitudes resembles that seen in the
conventional solution.

One can also renormalize after expansion by an integer amount,
$L$, greater than 2. In this case the discussion proceeds
exactly as above but with $a_{\rm max}=L^{(3+n)/2}$, while all terms
containing $\comb({\bf k}/2^n)$ become $\comb({\bf k}/L^n)$. 

The normalization of the mass-fluctuation power spectrum 
may be represented by the variance of the mass in a randomly
placed sphere of radius $R$, 
\beq
	\sigma^2(R)=\sum_{\bf k} 
	|\tilde\delta({\bf k})|^2\widetilde{W}^2(kR),
			\label{eq:c3} 
\eeq
where $\widetilde{W}(r)$ is the Fourier transform of a spherical
top-hat window. Since a particular value of ${\bf k}$ appears in
only one of the iterations of the sum over $j$ in
equation~(\ref{eq:c1}), we can write equation~(\ref{eq:c3}) as 
\beq
	\sigma^2(R)=\sum_{j=0}^{J-1} a_{\rm max}^{2j}
	\sigma_0^2(L^jR),\label{eq:c4}
\eeq
where
\beq
	\sigma_0^2(R)=\sum_{\bf k} A^2({\bf k})
	\prod({\bf  k})\widetilde{W}^2(kR)\label{eq:c5}
\eeq
is the mass variance input each new
iteration. We may approximate the sum over $\bf k$ in
equation~(\ref{eq:c5}) as an
integral in the usual way (with the restriction that $n>-3$). The
discreteness effects here will be more important than is usual
in making such an approximation. This is because of the small number
of waves, the fact that the upper $k$ cut at $k_c$ does not translate
into a smooth upper limit to $|{\bf k}|$, and the inclusion of the
roughly 1/8 of waves per unit volume of $k$ space which are missing in
$A$ in equation~(\ref{eq:c2}) because of the factor 
$1-\prod({\bf k}/2)$. Assuming 
that the top-hat filter $\widetilde{W}(kR)$ approximates a sharp
cutoff at $k=1/R$, we find
\beq
	\sigma_0^2(R)\ap [\min(1/R,k_c)]^{n+3}\quad
	\hbox{for $R \la 1$,\quad 0 otherwise};\label{eq:c6}
\eeq 
and hence
\beq
	\sigma^2(R)\ap R^{-(n+3)}\Bigl( 
	\lceil\log_L (k_c/2\pi)\rceil+ (a_{\rm max}^2-1)^{-1}\Bigr).
		\label{eq:c7}
\eeq 
Equation~(\ref{eq:c7}) is valid for $L^{-J} < R\la 1/k_c$. For larger $R$
there is a slow logarithmic departure from $R^{-(n+3)}$. The
first term represents the waves from renormalizations which have
some wavelengths of size comparable to the sphere size, the
second the cumulative effect of waves from renormalizations which
are much larger than the sphere size.  

\section{Parameters in the Numerical Solutions}

Units for the renormalization computation are chosen so the
comoving width of the cube is 
$r=1$, the initial value of the expansion parameter is $a=1$, the
gravitational constant is $G=1$, and the particle mass is $m=1/N$ for
$N$ particles in the cube. In these units the initial mean mass
density is $\rho _i=1$ and the initial time is $t_i=(6\pi
)^{-1/2}$. The pairwise particle force is that of the interaction between
``softened'' particles with density 
profile $\rho\propto(1-r/c)$ at $r\leq c$ 
and $\rho =0$ at $r>c$,  where
$c$ is an adjustable softening length. This is the standard
``S2'' particle shape of Hockney \& Eastwood~(1981). With this
definition of $c$, the peak in the force law is at $0.78c$. The
particle separation at the peak is 
close to that for a standard Plummer force, $r/(r^2+c^2)^{3/2}$,
which is at $c/\surd2$, but the transition between the linear form 
at $r\ll c$ and the inverse square form at $r\gg c$ is sharper
than for the Plummer force. The softening, or ``cutoff'', length
$c$ is held constant in comoving coordinates. 

Inter-particle forces in the periodic simulation cube are computed
using the AP$^3$M    
technique (Couchman~1991) and particle positions and
velocities are updated using time-centered leapfrog. The expansion
following each renormalization, which corresponds to a time range from
$t_i=(6\pi)^{-1/2}$ 
to $t_f=a_{\rm max}^{3/2}\, t_i$, is integrated in 400 equal time
steps. The results obtained using half this number of time-steps are
statistically indistinguishable from those presented below. 

After each renormalization, new modes are added by shifting the
particle velocities and positions in real space using the appropriate
sinusoidal wave displacements. Since we are adding relatively few
waves at each iteration (eq. [\ref{eq:c2}]) this does not
significantly increase the computational effort. The same method is
used to generate initial conditions for the conventional simulations,
but at considerably greater computational expense.

We define the normalization of the waves added at 
each iteration (set by $\beta$
in eq. [\ref{eq:c2}]) by $\sigma_0(R=0)$ using
equation~(\ref{eq:c5}), and we refer to this input parameter simply as
$\sigma_0$. This equals the rms fluctuation added at each iteration on
scales much smaller than $2\pi/k_c$ (eq.~[\ref{eq:c2}]). 

The convergence of the renormalization method 
is rapid, and typically after six iterations 
there is little scatter from iteration to 
iteration. To assure elimination of transients we evolve through
at least ten iterations prior to 
saving the first realization. In the realizations, particle
positions and velocities are saved at the end of the integration
step 
(immediately before renormalization), at every fifth iteration, to
accumulate five realizations. The five iterations through the
renormalization loop between saved realizations serve to
suppresses correlations among realizations. Error flags
shown in the next section are
the full spread of values from the five realizations. In the
numerical results the length unit for particle separations $r$ is
the width of the cube, and particle velocities use the proper length
$ra_{\rm max}$ and the proper time unit for the computation,
where the expansion time is $a_{\rm max}^{3/2}(6\pi )^{-1/6}$. 

We choose for our standard renormalization solution the
parameters  
\beq
	n = -1, \qquad N = 64^3, \qquad c = 0.001, \qquad\sigma_0 = 0.1.
		\label{eq:parameters}
\eeq
The power law index $n$ gives a reasonably close fit to the shape
of the galaxy two-point correlation function. 

The comparison solution, from a conventional N-body computation,
uses the first three parameters in
equation~(\ref{eq:parameters}). In an 
effort to make the comparison and standard renormalization solutions
as comparable as possible, in one of the conventional
realizations we use
the phases of the Fourier components that were applied at each
renormalization step for the corresponding set of wavenumbers. In the
conventional simulation, with $64^3$ particles, wavemodes up to 32
times the fundamental in the cube are used in the initial
conditions. Matching phases from
the renormalization computation involves saving the phases from the
last 6 iterations (since $2^{6-1}=32$) of the standard run. New random
phases are introduced for the Fourier components that were not applied
in any of these renormalization steps. The result is that one can see
some similarity of features in maps of particle positions in the
renormalization and conventional realizations. This is seen
in Figure~\ref{fig1}, which compares maps of particle positions
in a renormalization realization and in a ``phase-matched''
conventional realization. The statistical results in \S 4 are
averages across this ``phase-matched'' conventional realization and
four other conventional realizations initialized with
independently chosen random phases. 

The initial density fluctuations in the conventional simulation are
applied by the distortion of a distribution of particles initially at
the vertices of a cubic lattice. The net expansion factor and the
amplitude of the applied density fluctuations are set by the following
considerations.  In the numerical solution evolved from conventional
initial conditions the mass variance in spheres arising from a
linear input spectrum $\vert\delta({\bf k})\vert^2 =\beta^2 k^n$ is
$\sigma^2(R)\propto R^{-(n+3)}$, where the implied constant of
proportionality is the same as that in equation~(\ref{eq:c7}).  Thus
we must increase the amplitude of the waves input to the conventional
solution by a factor of $\Bigl(\lceil\log_L (k_c/2\pi)\rceil+ (a_{\rm
max}^2-1)^{-1}\Bigr)^{1/2}$ which, for $L=2$ and $k_c/2\pi=\surd22$, is a
factor of approximately 1.8. Note, that there are several
approximations involved in the derivation of equation~(\ref{eq:c7})
which weaken its predictive power and, furthermore, we cannot expect
accurate matching of the two types of initial conditions over the full
range of scales in the simulations. Enhancing the amplitude by the
factor indicated nonetheless gives a good match between the final
correlation amplitudes of the renormalized and conventional solutions.
 
In the standard renormalization solution the input fluctuation at
each iteration is $\sigma_0=0.1$. After four further
renormalisations these waves 
would dominate fluctuations on the scale of the mean interparticle
spacing and contribute an rms fluctuation, in linear theory,
amounting to $0.1\times2^4=1.6$. (We have ignored the
contribution from waves input five 
renormalisations ago as these all now have wavenumbers at or beyond
the Nyquist frequency.) The total rms fluctuation amplitude on
this scale from all iterations would be of order 1.8 times greater
(eq.~[\ref{eq:c7}]), giving a total linear rms fluctuation of
roughly three.  In order that all waves in the initial conditions
for the conventional solution are in the linear regime we have
started the conventional simulation with $a_i=0.1$ with a
corresponding decrease of a factor of 
ten in the input wave amplitudes. The rms fluctuations on the smallest
scales at the start of the conventional simulation are thus of order
0.3. A test with an initial amplitude 5 times lower (with 
expansion starting at $a_i=0.02$) yields results that are
statistically indistinguishable.

The time interval corresponding to expansion from $a=0.1$ to
$a_{\rm max}=2$ in the comparison solution is integrated using
time-centred leapfrog, as 
before, with 918 equal steps. Each timestep in this case corresponds
to 2/3 of that for the renormalization iterations. A shorter timestep
ensures that the rapid expansion at the start of the
conventional simulation is accurately followed. A test with twice as
many timesteps gives the same statistical results.

At relatively large separations we get good statistical reliability of
the estimates of the two-point position correlation function and the
one- and two-point velocity statistics by using a random sample of a
fraction $f=0.1$ of the particles. At separations $r<0.05$ the
statistics are based on all particle pairs except
where otherwise noted.

\subsection{Results}

In the maps in Figure~\ref{fig1} the length scale has been adjusted so that
the clustering length $r_o$ agrees with the galaxy clustering
length, and the selection function is a rough approximation to that of the
Las Campanas redshift survey~(Schectman {\it et al.}~1996; Lin {\it et
al.}~1996). Parts (a) and (b) show a slice in the 
standard renormalization solution, and parts (c) and (d) show the
same slice in the comparison conventional solution in which 
wavenumbers in common with those applied at the renormalization
steps have the same phases (the ``phase-matched'' pair). 
In (a) and
(c) the radial variable is the distance, and in (b) and (d) it is
the redshift. Because many of the phases of Fourier components for
low wavenumbers
are the same one can see similarities in the clustering pattern
in the renormalization and conventional realizations. 

In Figure~\ref{fig1} and other maps of the particle distributions the
low density regions between the prominent mass concentrations are more
nearly empty in the renormalization  
case. In maps of all particle
positions in thin parallel slices one can see in the low density
regions remnants of the perturbed lattice in the initial conditions
for the conventional solution. They do not appear in the
renormalization solution, of course. The ``walls'' in the redshift
maps are less prominent in the renormalization solution.  This may be
a result of the stronger small-scale clustering and weaker large-scale
motions in the renormalization solution. On scales $\ga r_o$   
infall velocities $|v(r)|$, as measured by the mean pairwise 
velocity, are systematically lower in the 
renormalization solution. This will tend to de-emphasize the 
sharpness 
of walls in redshift-space in the renormalization solution.  The 
fractional difference of $|v(r)|$ in renormalization and 
conventional solutions is about 10\% at the correlation 
length (0.013 times the
box width) rising to 20\% at 0.1 times the box width. The 
differences thus are not large but perhaps contribute to the 
difference in appearance of the maps.

Figure~\ref{fig2} shows how parameter changes affect the two-point position
correlation function $\xi(r)$ in the renormalization solution.
The correlation functions are based on a sampling fraction of 0.1 of
the particles, and the error bars are the maximum scatter across
five realizations. The correlation function is defined by the
usual relation, $1+\xi = 
N_p/(n\delta V)$, where $N_p$ is the mean number of neighbors of a
particle in the distance range of the spherical shell with volume
$\delta V$, and $n=N/V$ is the mean particle number density in
the box. There is no edge correction because space is
periodic. The total number of pairs is fixed, and the correlation
function therefore satisfies the usual constraint, 
$\int d^3r\, \xi (r)=0$ for $N\gg 1$, where the integral is over the box
volume.

The standard renormalization solution in Figure~\ref{fig2} is shown as 
triangles. In the solution plotted as asterisks the amplitude of the
applied density fluctuations is a factor of three larger than in
the standard solution. In a self-similar solution this is
equivalent to adjusting the length scale by the factor $r\rightarrow
3^{2/(3+n)}r$ (eq.~[\ref{eq:pert}]). Since $n=-1$, the particle
separations in Figure~\ref{fig2} have been scaled by a factor of three. The
gravitational interaction cutoff length in this solution is $c=0.003$,
so this length appears in the plot at separation $0.001$, at the
right-hand arrow. The ratio of clustering length $r_o$ to inverse
square cutoff length, $c$, thus is the same in the standard solution
(triangles) and this solution plotted as asterisks. The
difference between these solutions at one third the cutoff length is
an indication of sensitivity to parameters in the
renormalization method. The close agreement at larger separations
is a significant indication that we have a good approximation
to the 
self-similar solution. In the solution plotted as squares the applied
amplitude is three times the standard solution and the cutoff length,
$c$, is the same so $c$ appears in Figure~\ref{fig2} at separation 0.00033, at
the left-hand arrow. This solution also is in satisfactory agreement
with the standard one. Finally, the circles show a solution with
particle number three times the standard case. As in the other
solutions the correlation function is the average across five
realizations each separated by five renormalization steps.  We see
that changing $N$ has very little effect on the two-point correlation
function. 

Efstathiou and Eastwood (1981) introduced another useful test of
self-similarity, based on the relation between the two-point position
correlation function $\xi (r)$ and the mean relative peculiar
velocity $v(r)$ of particle pairs at separation $r$,
\beq
	{\partial\xi\over\partial t} =-
	{1\over r^2a}{\partial\over\partial r}r^2
	(1+\xi )v(r) =
	-\alpha {r\over t}{\partial\xi\over\partial r}.
		\label{eq:pairconservation}
\eeq
The first part expresses conservation of particle
pairs. The second part follows from the scaling relation in 
equations~(\ref{eq:alpha}) and (\ref{eq:alpha(n)}). In a
self-similar solution $v(r)$ derived from $\xi (r)$ using the second
part of equation~(\ref{eq:pairconservation}) agrees with the
mean relative peculiar velocity of particle pairs at separation
$r$. The second part of equation~(\ref{eq:pairconservation}),
after multiplication by $r^2$ and integration over $r$ (by parts
on the right hand side), gives 
\beq
      v(r)= {\alpha\, a\over t} {r\over 1+\xi (r)}\Biggl\{\xi (r)
          -\int_V\xi\,dV/V\Biggr\},\label{eq:intpaircons}
\eeq
where $V$ is the volume of a sphere of radius $r$.
This is a convenient form with which to test for self-similar
behavior: $\xi$ is measured in the 
straightforward manner described above, whilst 
$\int_V\xi\,dV$ is obtained directly from the mean count of
neighbors within distance r of a particle.

In the bottom panels in Figures~\ref{fig3} to~\ref{fig5} the solid
line is the result of 
predicting $v(r)$ from the measured $\xi(r)$ using
equation~(\ref{eq:intpaircons}). The open circles in the lower panels
are the means $\lb v_{ij}\rb$ of the relative peculiar velocities of
particle pairs $(ij)$, and the error flags are the scatter of the
means across the five realizations, in both renormalization and
conventional solutions.  The error flags for the $\lb v _{ij}\rb$ at
the smallest separations plotted are larger in the conventional
solution because the rms scatter in relative velocities of
particles is larger and the number of pairs is smaller.

In the standard renormalization solution shown in Figure~\ref{fig3} the
mean relative velocities of particle pairs is quite close to what
is expected in a self-similar solution at separations smaller than
about $2r_o$, while at larger separations there is a significant 
systematic difference. The same is true of the 
renormalization solution for $n=0$ in Figure~\ref{fig4}. This difference
between $\lb v_{ij}\rb$ and $v(r)$ at large $r$ might be a result of
the disturbance of the system by renormalization at every
factor $2^{(3+n)/2}$ expansion, and we might expect that this 
disturbance is largely forgotten on smaller scales, consistent
with the success of the scaling test. The discrepancy goes the
other way in the conventional solution in 
Figure~\ref{fig5}, again as one might have anticipated: the solution is a 
good approximation on mildly nonlinear scales but does less well
in the deeply nonlinear sector where there may be incomplete 
suppression of transients from the initial conditions. In the
conventional numerical self-similar solution (from S. White) used
by Jain~(1997), with particle number $N=100^3$ and power law
index $n = -1$, the 
departure from self-similar behavior is less significant than in
our conventional solution in Figure~\ref{fig5} although the error bars
are significantly bigger in Jain's Figure~2. 

The solid lines in the upper panels of Figures~\ref{fig3}
to~\ref{fig5} are the power law,
\beq 
	\xi\propto r^{-\gamma (n)},\qquad 
		\gamma (n)= (9+3n)/(5+n).\label{eq:gamma}
\eeq
This applies in the self-similar solution when the time scale for the 
evolution of the mean clustering (measured in physical length
units) is much longer than the Hubble time. In this case the mean
relative peculiar velocity is the negative of the Hubble
relation, which is shown in the dashed curve in the lower panels of the
figures. All solutions here and in Jain (1997) agree that there
is significant infall in 
physical length units at separations comparable to the clustering
length $r_o$. This infall makes the logarithmic
slope of $\xi (r)$ steeper than $\gamma (n)$. In the standard
solution with 
$n=-1$ the result is not far from the power law shape of the 
galaxy two-point correlation function. The departure from a
power law is more prominent in Jain's (1997) solution, 
perhaps
because the larger particle number allows the solution to reach
larger values of $\xi$. At $n=0$ the slope of $\xi (r)$ in our
solution is steeper than the galaxy function at $r\sim r_o$, and
the departure from a power law more prominent.

Figures~\ref{fig6} and~\ref{fig7} compare the two-point
correlation functions in 
the numerical solutions to the method of Hamilton~{\it et~al.}
(1991) for interpolation between equation~(\ref{eq:gamma}) 
on small scales and linear perturbation theory on large scales.
The dashed curves are the fitting function for the Hamilton~{\it et~al.}
method from Peacock \& Dodds (1996), and the solid curves are the
fitting function of Jain, Mo, \& White (1995). The feature in the
latter at $r\sim r_o$ is much less prominent in the integral 
$\int _0^rr^2\xi (r)dr$ considered by Jain {\it et al.}, but the
Peacock \& Dodds form does better fit our renormalization solutions.
This feature aside, there is very close consistency with
the renormalization solutions at $n=-1$ and $n=0$ and with the
conventional solutions for correlation amplitudes around unity.

It might be mentioned that the Hamilton {\it et~al.} method
assumes statistically stable small-scale clustering, as
reflected in equation~(\ref{eq:gamma}). Stability is difficult to
demonstrate on theoretical or numerical grounds (Ruamsuwan \&\
Fry 1992; Jain 1997); the striking success of the 
interpolation method offers some support for stability.

Figure~\ref{fig8} shows second moments of the relative peculiar
velocities of 
particle pairs as a function of their separation, in the standard
renormalization and conventional solutions with $n=-1$. The rms
dispersion $\sigma _r$ in the component along the line
connecting the particles is computed relative to the mean value
$\lb v_{ij}\rb$. The transverse component $\sigma _t$ is
normalized to one direction orthogonal to the line
connecting the particle pair, so $\sigma _r=\sigma _t$ if the 
velocities are isotropic relative to the mean. The 
renormalization and conventional 
solutions agree that $\sigma _r$ is systematically greater than
$\sigma _t$ at $r>c$. The velocity dispersion is distinctly
larger in the conventional solution, 
and the difference grows with decreasing separation to a factor
of two at the smallest separation plotted (which is well within
the force cutoff length $c$). If the mass autocorrelation
function at small scales varied as 
$\xi\propto r^{-\gamma}$, and the clustering on average were not 
evolving (in physical units), the relative velocity dispersion
would vary as  $\sigma\propto r^{1 -\gamma /2}$. There is not
enough range between $c$ and $r_o$ for a test, but the
renormalization solution does show the expected slow increase of
the relative velocity dispersion with increasing separation of
the particles at $r\la r_o$. 

Figure~\ref{fig9} shows distributions of the absolute values of
one Cartesian component of the relative proper velocity of
particle pairs. We compare the ``phase-matched'' pair of
conventional and renormalization realizations; the distributions
from the other four members of each 
of the ensembles for $n=-1$ have the same main features.
Consistent with Figure~\ref{fig8}, the renormalization solution
has a much narrower distribution at small separation and
a perceptibly narrower distribution at relatively large
separation. 
 
Figure~\ref{fig10} shows distributions of the counts of neighboring
particles within given distances of a particle. The abscissa is the
count $n$ of neighbors. The ordinate is the fraction of particles that
have $n$ or more neighbors within the given distance. The
histograms for the renormalization realizations
are plotted as solid lines and for the conventional
realizations as dotted lines. The ``phase-matched'' pair are
the heavy solid and dotted lines. If the richest concentration
within scale $r$ contained $n_x$ particles, and the next richest
concentration on this scale contained distinctly fewer, 
it would produce a shoulder or near level section at probability
$\sim n_x/64^3$ for $\sim n_x$ neighbors. A prominent
example is in the 
``phase-matched'' conventional realization at $r=0.005$. As
it happens, at this radius the distributions in the
``phase-matched'' pair are at the extreme high and low sides of
the realizations even though this pair was designed to have
similar space distributions. There
are too few realizations to decide whether the shoulder effect is
more common in the renormalization or conventional approach. 

Despite the scatter among realizations there is a clear 
trend in Figure~\ref{fig10}: the renormalization solution tends 
to have less extreme mass concentrations at $r\ga 0.005$ and
larger concentrations on smaller scales. At $r\la 0.002$ the
difference is seen in the second moment (Fig.~[\ref{fig6}]), but
at larger separations it
has little effect on $\xi (r)$ or on the third moment, as
discussed next. The rms scatter in the relative gravitational
acceleration of particle pairs depends on the fourth moment, and
the differences in distributions of neighbours likely accounts for the
systematic difference in relative velocity dispersions. At large
separations the smaller velocity dispersion in the
renormalization solution would be expected from the less common 
occurrence of strong mass concentrations. At small separations the
more numerous close 
pairs in the renormalization solution likely are in a
clustering hierarchy that extends to smaller scales, whereas
close pairs in the conventional solution are more likely to be
accidentals moving with larger relative velocities in larger
clumps. 

Figure~\ref{fig11} shows a commonly applied measure of the skewness of
the mass distribution in randomly placed cells. Because there is
some potential for confusion we remind the reader of the
following results. A stationary and isotropic random point
process with mean particle number density $n$ may be
characterized by its $N$-point correlation functions. The second
and third are defined by the joint probabilities of finding
particles in two and three disjoint volume elements:
\beq
	dP_{12} = n^2[1+\xi (12)]dV_1dV_2,\label{eq:xi}
\eeq
\beq
	dP_{123} = n^3[1 + \xi (12)+ \xi (23)+ \xi (31) +\zeta (123)]
	dV_1dV_2dV_3.\label{eq:ze}
\eeq
The argument of the two-point function is the distance between
the two volume elements. The arguments of the three-point
function are 
the three sides of the triangle defined by the three volume
elements. The mass in a randomly placed sphere of radius $r$ and
volume $V$ is $M=mN$, where $m$ is the particle mass. The
expectation value is $\bar M =m\bar N$, where $\bar N=nV$. The
mass density contrast in the sphere is
\beq
	\delta = M/\bar M - 1.
\eeq
The ratio of the third central moment to the square of
the second central moment of the probability distribution in the
mass contrast is  
\begin{eqnarray}
  S_3  &=& {\langle\delta ^3\rangle\over\langle\delta ^2\rangle ^2} 
     = {\bar N\lb (N-\bar N)^3\rb\over [\lb (N-\bar N)^2\rb]^2}\nonumber \\
\noalign{\vskip -5pt}
	& &  \label{eq:S3}\\
\noalign{\vskip -5pt}
	&=& {\int\zeta (123)\, d^3V/V^3 + 
	3\bar N^{-1}\int \xi (12)\, d^2V/V^2 +\bar N^{-2}\over
	[\int\xi (12)\,d^2V/V^2 + \bar N^{-1}]^2}. \nonumber
\end{eqnarray}
The integrals are over the sphere volume $V$. 
If the points were tracers of the distribution of an underlying
continuous fluid with mass density $\rho ({\bf r})$, such that
a point is placed in the volume element $dV$ at position {\bf r}
with probability \beq dP\propto\rho ({\bf r})dV,\eeq then 
the shot or discreteness noise terms in equation~(\ref{eq:S3})
would be a consequence of the limited sampling of the mass
distribution. In this case the shot noise terms should be
removed. This is done in the expression  
\beq
	S_3^\prime = {\int\zeta (123)\, d^3V/V^3\over
	[\int\xi (12)\,d^2V/V^2\, ]^2}
	={\bar N[\lb (N-\bar N)^3\rb - 3\lb (N-\bar N)^2\rb
	+ 2\bar N]\over [\lb (N-\bar N)^2\rb -\bar N]^2}.
			\label{eq:S3f}
\eeq

One can define mass correlation functions by replacing each 
particle with a sphere of volume $\delta V$ and internal density
$m/\delta V$. In the limit $\delta V\rightarrow 0$ the mass 
two-point or autocorrelation function and the mass three-point
function are 
\begin{eqnarray}
  \xi _\rho (12) &=& \xi (12) + \delta (12)/n,\nonumber \\
\noalign{\vskip -5pt}
	& &  \label{eq:xirho}\\
\noalign{\vskip -5pt}
  \zeta _\rho (123)&=& \zeta (123) + [\delta (12)\xi (23) +
	\delta (23)\xi (31) +\delta (31)\xi (12)]/n + 
	\delta (12)\delta (23)/n^2, \nonumber
\end{eqnarray}
where the position correlation functions $\xi$ and $\zeta$ are
defined in equations~(\ref{eq:xi}) and~(\ref{eq:ze}). 
When $\xi$ and $\zeta$ are replaced by $\xi _\rho$ and 
$\zeta _\rho$ in the ratio of integrals in
equation~(\ref{eq:S3f}) the Dirac delta functions in
equation~(\ref{eq:xirho}) produce the
extra terms in equation~(\ref{eq:S3}).

In numerical N-body solutions such as the ones used in this study
the mass distribution actually is that of the particles, and
therefore equation~(\ref{eq:S3}) (or the ratio of integrals in 
eq.~[\ref{eq:S3f}] over the mass functions in eq.~[\ref{eq:xirho}])
is the ratio of moments
of the mass distribution that figures in the dynamics. The
difference between equations~(\ref{eq:S3}) and~(\ref{eq:S3f})
(where $\xi$ and $\zeta$ are the position correlation functions 
in eqs.~[(\ref{eq:xi}] and~[\ref{eq:ze}]) is 
a useful measure of the relative
contributions to $S_3$ by shot noise and the clustering of 
particles. In a good approximation to a self-similar solution the
shot noise is subdominant, of course.

Another measure of the importance of shot noise is provided by
the two ratios  
\beq
	\lb (N-\bar N)^2\rb /\bar N, \qquad 
	\lb (N-\bar N)^3\rb /[3\lb(N-\bar N)^2\rb ] .\label{eq:shot}
\eeq
When $n$ is sufficiently large (and the particle mass
correspondingly small), as in a good self-similar solution, these
ratios are much larger than 
unity and equations~(\ref{eq:S3}) and~(\ref{eq:S3f}) are
equivalent. 

The statistics in Figure~\ref{fig11} are derived from the
``phase-matched'' pair of realizations for $n=-1$.
The ratios in equation~(\ref{eq:shot}) are plotted as
the dotted curves. 
They show that the shot noise contribution to the skewness and
variance is reasonably small at cell radius $r$ equal to the
clustering length $r_o$, but shot noise is dominant
in the $n=-1$ case at $r\la 0.1r_o$. (The situation is a little
better at $n=0$ because $\xi (r)$ is larger.) This is our most
vivid illustration of the limited dynamical range in length scale 
available for N-body simulations of the nonlinear sector of
clustering solutions even when $N^{1/3}$, which sets the
dynamical range, is within a factor of four of 
what is now feasible. 

The solid lines in Figure~\ref{fig11} are the ratios $S_3$ of mass moments
(eq.~[\ref{eq:S3}]) for $n=-1$. To illustrate the contribution of
the shot noise terms to the mass moments we show as the dashed
lines $S_3^\prime$ based on the reduced correlation functions 
(eq.~[\ref{eq:S3f}]). Colombi, Bouchet, \&\
Hernquist (1996) study $S_3^\prime$ in their conventional
self-similar solution. Their results are in reasonably close
agreement with $S_3^\prime$ from our conventional
solution at $r\la 0.001$, but they do not find the rapid
increase in noise at smaller separation.

The comparison of solid and dashed lines in Figure~\ref{fig11}
shows that the 
shot noise contribution to $S_3$ is appreciable but not dominant
at $r=r_o$, and the shot noise contribution diverges at
$r=0.1r_o$. This agrees with the measures of shot
noise in equation~(\ref{eq:shot}). 

At large sphere radius shot noise is unimportant and
second-order perturbation theory should apply. Here 
Bernardeau (1994a,b) finds
\beq S_3 = 34/7 - (3 + n),\label{eq:Bern}\eeq
for Gaussian initial density fluctuations, and
\beq S_3^{\rm Zel} = 4 - (3 + n), \label{eq:B-Zel}\eeq
for the Zel'dovich approximation. 
Equation~(\ref{eq:Bern}) is plotted as the horizontal dotted lines
in Figure~\ref{fig11}. The upturn in $S_3$ at $r\sim 0.1$ 
in the numerical solutions is of doubtful significance because
the sphere radius 
approaches the box size. At $r_o\la r\la 0.1$ the conventional
solution is somewhat closer to perturbation theory with Gaussian
initial conditions, consistent
with the somewhat better large-scale performance of the
conventional solution in the scaling test in Figures~\ref{fig3}
to~\ref{fig5}. In the
renormalization solution, density fluctuations are applied by the
Zel'dovich approximation, and the lower skewness in
equation~(\ref{eq:B-Zel}) might be expected to persist through 
several iterations. This may contribute to the 
lower value of $S_3$ for the renormalization solution. 

At $r\la c=0.001$ the correlation functions in the
renormalization and conventional solutions are quite different
(Fig.~[\ref{fig6}]). This difference in $\xi (r)$ is not reflected
in $S_3$, because the central mass moments are dominated by
discreteness noise and the mass moments in $S_3$ consequently are
at the shot noise limit.

At $c\la r\la r_o$ the skewness of the mass distribution can be
compared to that of the galaxy distribution. Useful approximations to
the galaxy correlation functions at $r\la r_o$ are \beq \xi =
(r_o/r)^\gamma, \qquad \zeta = Q(\xi _{12}\xi _{23} + \xi _{23}\xi
_{31} + \xi _{31}\xi _{12} ), \label{eq:zeta}\eeq where \beq
\gamma = 1.8, \qquad 
Q\simeq 1.\eeq A Monte Carlo integration over the three-point function
gives $S_3 = 3.1Q$ for $\gamma = 1.8$ and $S_3 = 3.0Q$ for $\gamma =
2$. A logarithmic slope $\gamma =-2$ is appropriate for the two-point
correlation function of the $n=-1$ renormalization solution over the
range of scales below $r_o$ for which the measured $S_3$ is not
completely dominated by shot noise ($10^{-3}\la r\la 10^{-2}$). The
value of $S_3$ in the numerical solutions is in line with these
numbers, although the comparison is not very sharply defined
because of the large shot noise component in the solutions. The
comparison of the self-similar solution and the galaxy
distribution is of limited significance in any case, of course, 
for if the expansion of the universe is scale-invariant 
(Einstein-de~Sitter) galaxies do not trace mass and if galaxies 
are useful mass tracers the expansion is not scale-invariant. 

\section{Discussion}

N-body approximations to self-similar clustering are severely
limited by shot noise. In our standard solution for $n=-1$ 
there is only a factor of ten difference between the nonlinear
clustering length $r_o$ and the sphere radius at which shot noise
dominates the mass moments (Fig.~[\ref{fig11}]). In this  
aspect our numerical solutions are quite unrealistic  
approximations to self-similar solutions at  
$r\la 0.1r_o\sim 0.001\sim c$. The parameters in 
equation~(\ref{eq:parameters}) were chosen 
so that shot noise dominates roughly at 
the force law cutoff $c$.\footnote{This
situation is not likely to be improved by  
increasing $r_o$. In our standard renormalization solution (with 
$r_o=0.013$) the rms fluctuation in counts in spherical cells 
is $\delta N/N=1$ for sphere diameter $2r=0.04$. For the primeval
mass fluctuation spectrum $P(k)\propto k^{-1}$
this would scale to rms fluctuation $\delta N/N=0.1$ at diameter
$2r=0.4$, if the count were not fixed in the box that has roughly 
twice this width. If $r_o$ were significantly increased 
the two-standard-deviation mass fluctuations on the scale of half 
the box width would be mildly nonlinear and seriously constrained 
by the fixed number of particles in the box.} 

In the numerical renormalization approach the value of $r_o$
changes by a 
factor of two between iterations. The factor of ten range of
scales between shot noise domination and 
nonlinear clustering at the end of the integration step 
accommodates this factor of two swing of $r_o$, but the
situation is at best marginal. The small-scale behavior of the 
renormalization solution thus must be treated with caution. 
We expect that the small-scale properties of a conventional solution
with the same parameters are 
even less secure because of the difficulty of establishing that
the solution truly approaches self-similar behavior. Thus we
suspect the pronounced difference of clustering
properties in our renormalization and conventional
solutions with $N=64^3$ (Fig.~[\ref{fig10}]) is mainly 
the fault of the latter. 

Relaxation is assured in the numerical renormalization method,
but at the price of a much poorer treatment
of initial conditions. The coarse population of initial Fourier
components in the renormalization approach is shown in
Figure~\ref{fig0}. For this reason we are inclined
to place greater trust in the properties of conventional
numerical solutions on mildly nonlinear scales. Our
interpretation is consistent with the mean relative velocity test
for self-similar behavior (Figs.~\ref{fig3} and~\ref{fig5}):
the renormalization solution does better on scales less than the
clustering length $r_o$ and the conventional solution does
better on larger scales.

Despite the shortcomings of the conventional and renormalization
methods the self-similar two-point mass autocorrelation function
seems to be quite accurately and reliably established 
at $\xi\la 100$. This is indicated by the excellent consistency 
of the renormalization and conventional solutions for $\xi (r)$. 
In particular, $\xi (r)$ in our renormalization solutions must be 
quite close to the conventional solutions used to find the
fitting functions for the Hamilton~{\it et~al.} interpolation
(Figs.~\ref{fig6} and~\ref{fig7}). As we have noted, conventional
and renormalization solutions are in overall good agreement with
the relative velocity test for self-similar evolution. 
Additional checks of the renormalization solution are the 
amplitude scaling (Fig.~\ref{fig2}) and the stability under
change of the particle number $N$. Indeed, the original results
at $N=1000$ are not much different from what we find at 
$N=3\times 64^3$. 

The relative velocity dispersion (Fig.~\ref{fig8}) and the
frequency distributions of relative velocities and cluster masses
(Figs.~\ref{fig9} and~\ref{fig10}) are more demanding,
and the comparison of renormalization and conventional solutions
is much less less satisfactory than for $\xi (r)$. Also
disturbing is the difference of appearance of the voids and walls
in the renormalization and conventional maps of particle positions 
(Fig.~\ref{fig1}).

The discrepancies between conventional and renormalization
solutions suggest that the numerical N-body predictions are
uncertain on issues that are observationally relevant
and important for theoretical analyses of self-similar evolution. 
Our understanding of these issues would be improved by using 
larger particle number $N$. An increase from 
the value in most solutions presented here, $N=64^3$, to
$N=128^3$ has already been done for the conventional
N-body method. An ensemble of five renormalization solutions
requires about seven times the computation for five conventional
solutions, which is feasible with present technology at
$N=128^3$. With all other parameters unchanged this would increase the 
mean number of neighbors at given comoving distance by a factor
of eight, increasing the ratio of clustering length to the radius
at shot noise dominance by the factor $2^{3/(3-\gamma )}\sim 5$. 
In addition to the exploration of differences between
conventional and renormalization solutions, the larger particle
number might allow a preliminary exploration of two questions. 
First, does the clustering hierarchy in the mass distribution, as
reflected in the hierarchy of N-point correlation functions (as
in eq.~[\ref{eq:zeta}]), persist to scales much smaller than the
clustering length $r_o$? An alternative is that 
merging produces monolithic massive halos with radii scaling with 
$r_o$, as assumed by 
Sheth \&\ Jain (1997). Second, are the 
distributions of relative velocities and  
positions consistent with statistically stable 
clustering on small scales, as assumed in
equation~(\ref{eq:gamma})? We hope to present results on these
issues from the analyses of renormalization and conventional
solutions with $N=128^3$ in due course.

\acknowledgments

We are grateful to Roman Scoccimarro
for discussions that substantially improved this paper. 
The work was supported in part at the University of Western
Ontario by NSERC of Canada, and at Princeton University by the US
National Science Foundation. HMPC thanks CITA for
hospitality whilst this work was completed.

\newpage
\bigskip

\figcaption[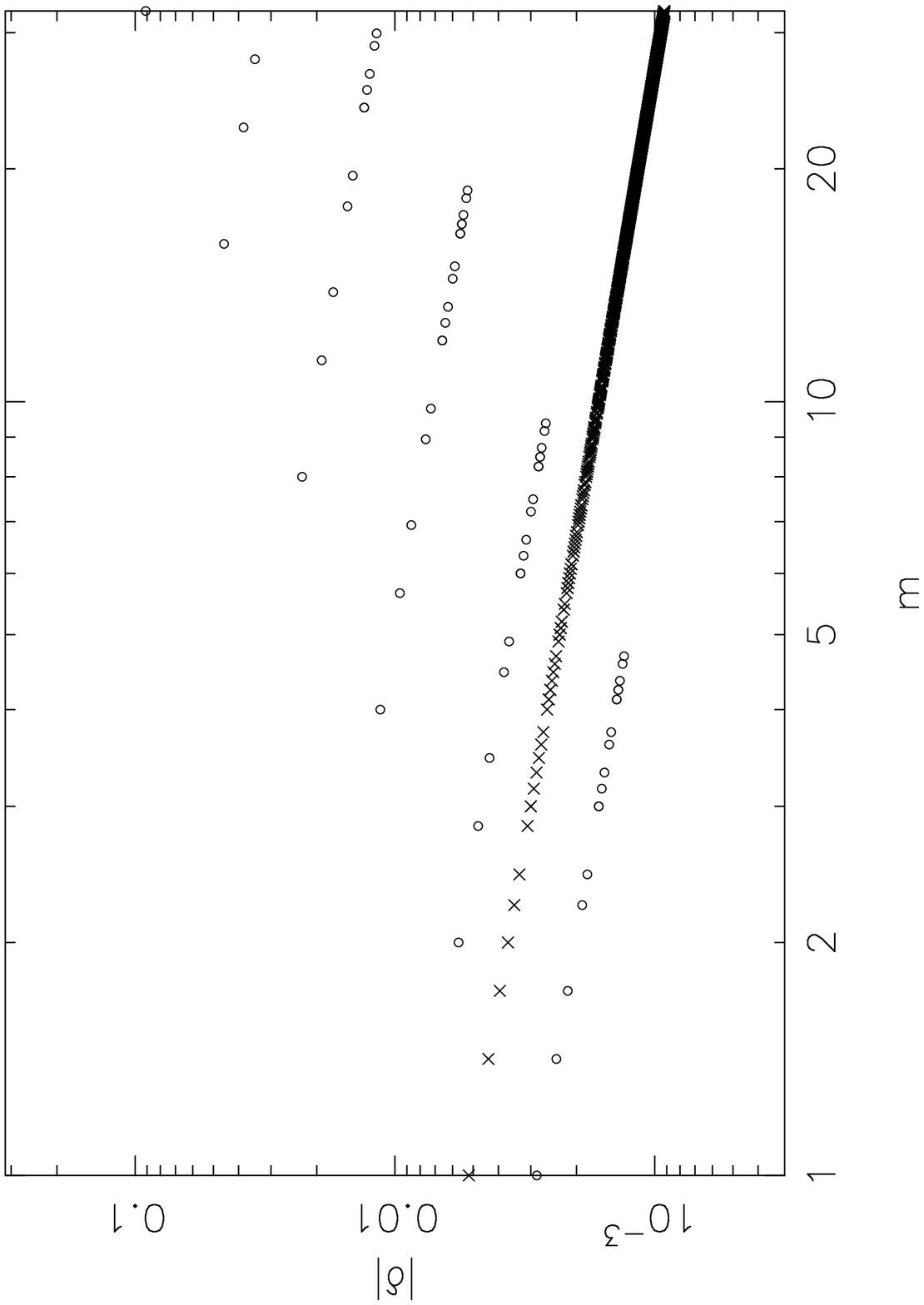]{Modulus of the applied Fourier components in
the renormalization (circles) and conventional (crosses) methods
(eqs.~[\ref{eq:c1}] and~[\ref{eq:c2}]) for $n=-1$. For the  
renormalization method 
we assume linear perturbation theory, so at each iteration wavenumbers
of previously applied components are doubled and the amplitudes
$|\delta _{\rm k}|$ are multiplied by the factor $a_{\rm
max}=2^{(3+n)/2}$. Wavenumbers are plotted in units of the fundamental
wavenumber in the box. The input amplitude corresponds to
$\sigma_0=0.1$ (eq.~[\ref{eq:c5}]). \label{fig0}}

\figcaption[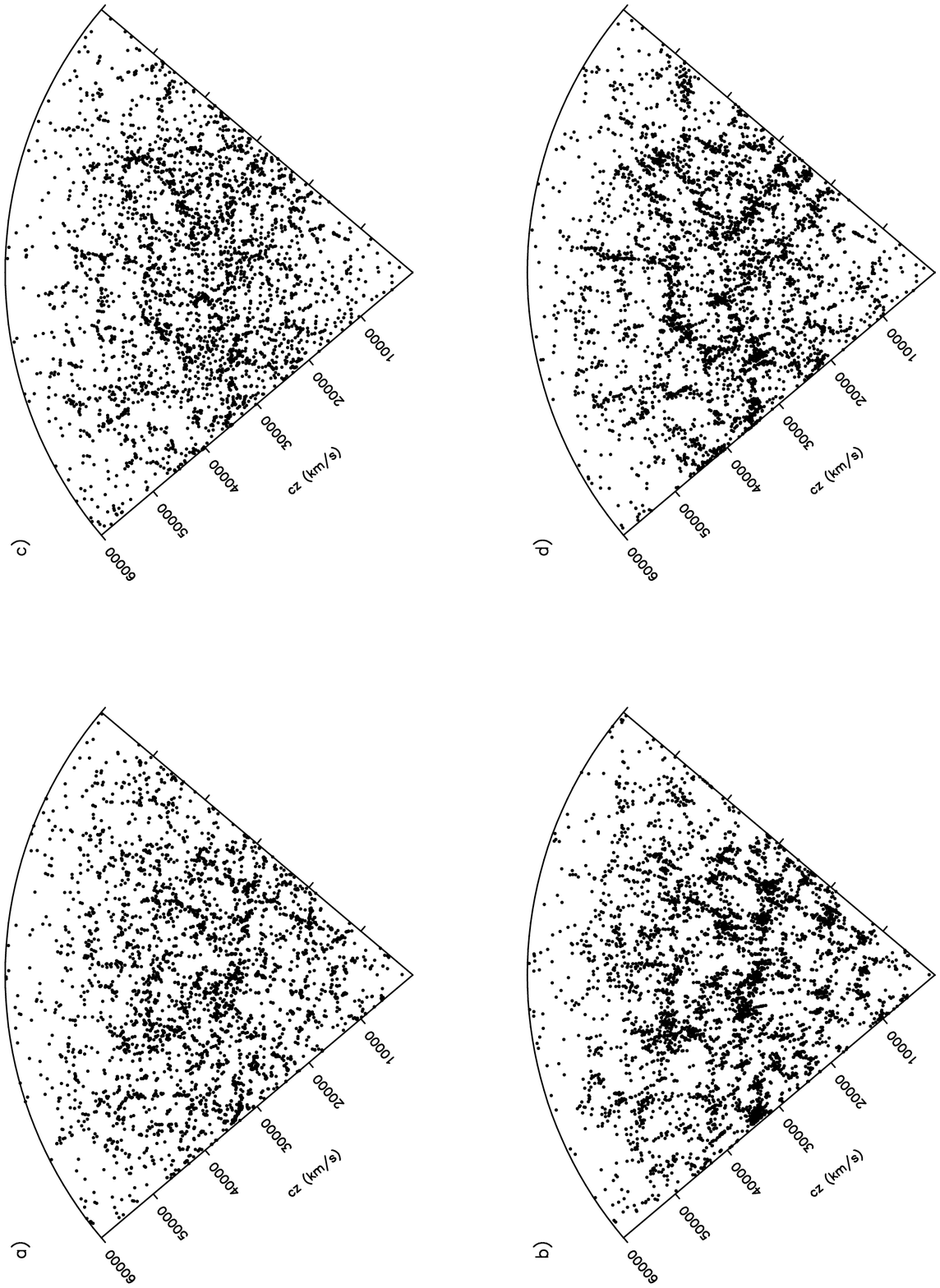]{Particle position maps in real (top) and
redshift (bottom) space in strips of width corresponding to 6~hrs of right
ascension at declination 32$^\circ$ and depth
1.5$^\circ$. The standard renormalization solution is on the left,
and the comparison conventional solution on the right. The length
scale has been adjusted to make the 
clustering length equal to $H_o r_o=540$ km~s$^{-1}$, close to
what is observed for galaxies. The fraction of particle
positions plotted as a function of distance approximates the
selection function of the Las Campanas redshift survey. \label{fig1}}

\figcaption[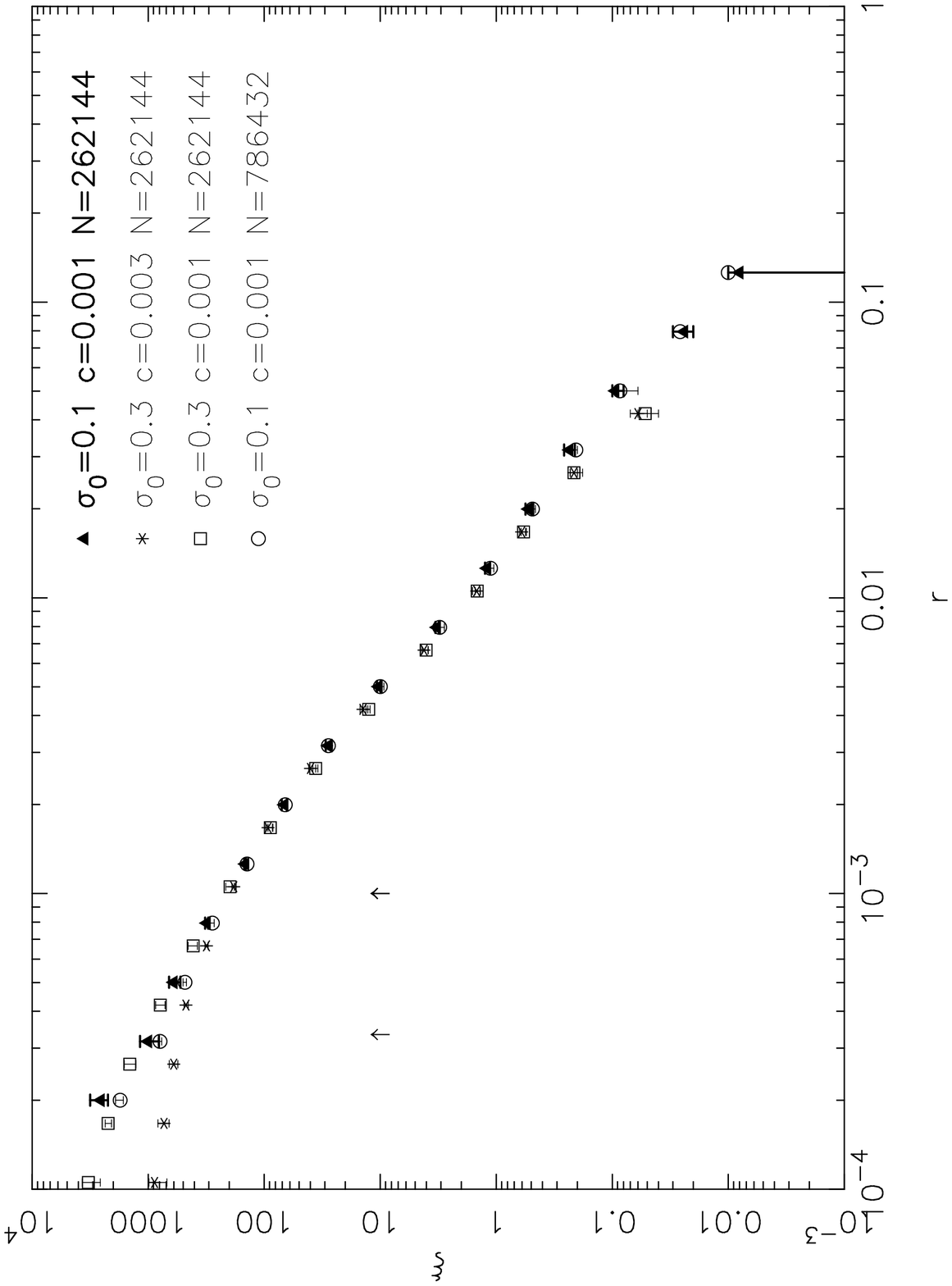]{Test of sensitivity of the position
two-point correlation function $\xi (r)$ to parameters in the
renormalization computation. The standard
renormalization solution is shown as
triangles, and the separation $r$ on the abscissa is plotted in
the units of this solution (where the box width is $r=1$).
Asterisks show the result of
increasing the amplitude $\delta$ of the applied perturbation and
the cutoff length $c$ of the gravitational interaction by factors
of three. Because $n=-1$, we compensate for the larger amplitude 
by scaling lengths by a factor of three (eq.~[\ref{eq:pert}]). 
The cutoff length thus is plotted at separation 0.001, at the
right-hand arrow, the same as the standard solution. In the
solution shown as squares the amplitude 
also is a factor of three larger than standard, so lengths have
been scaled by a factor of three, and the cutoff length $c=0.001$
appears at the left-hand arrow. The circles show the effect of
increasing the particle number by a factor of three. Here
$\delta$ is the same as the standard solution so the separations
are plotted at the coordinate values in the solution. \label{fig2}} 

\figcaption[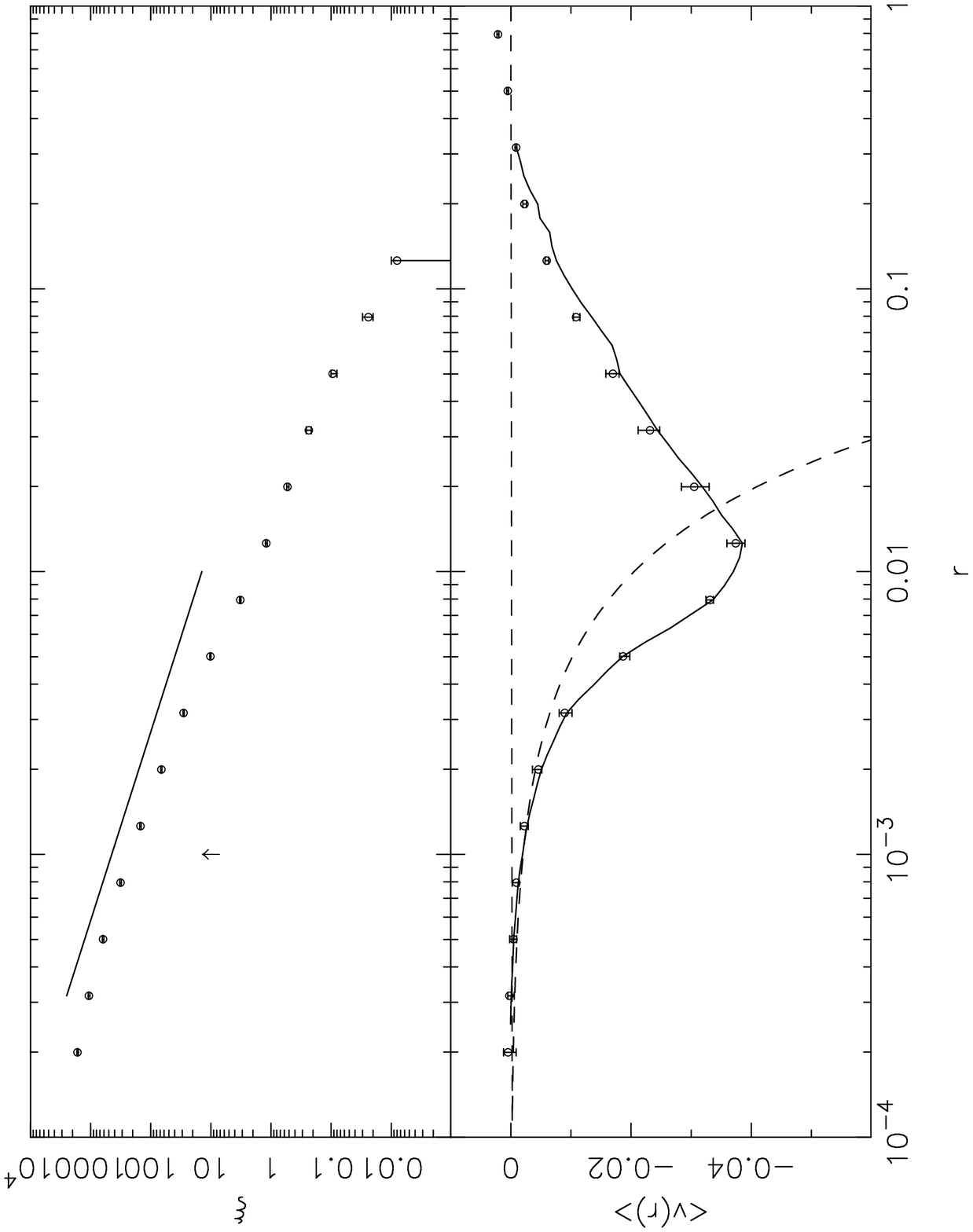]{Scaling test for the standard
renormalization solution ($n=-1$). The solid line in the upper
panel is the power law with index $\gamma =(9+3n)/(5+n)$
for statistically stable clustering, and the
circles in the upper panel are the two-point position correlation
function. The scatter across the five realizations is smaller
than the circles except at the three largest separations in the
plot. The arrow is the cutoff length $c$ for the
inverse square force law. The circles in the lower panel  
are means of the relative peculiar velocities of particle pairs, 
and the solid curve is the mean velocity derived from the scaling
law for the two-point correlation function,  
$\xi = \xi (r/t^\alpha)$ (eq.~[\ref{eq:intpaircons}]). The dashed
curve in the lower panel is   
$v=-r\dot a/a$, the peculiar relative velocity for physically
stable mean clustering. \label{fig3}}  

\figcaption[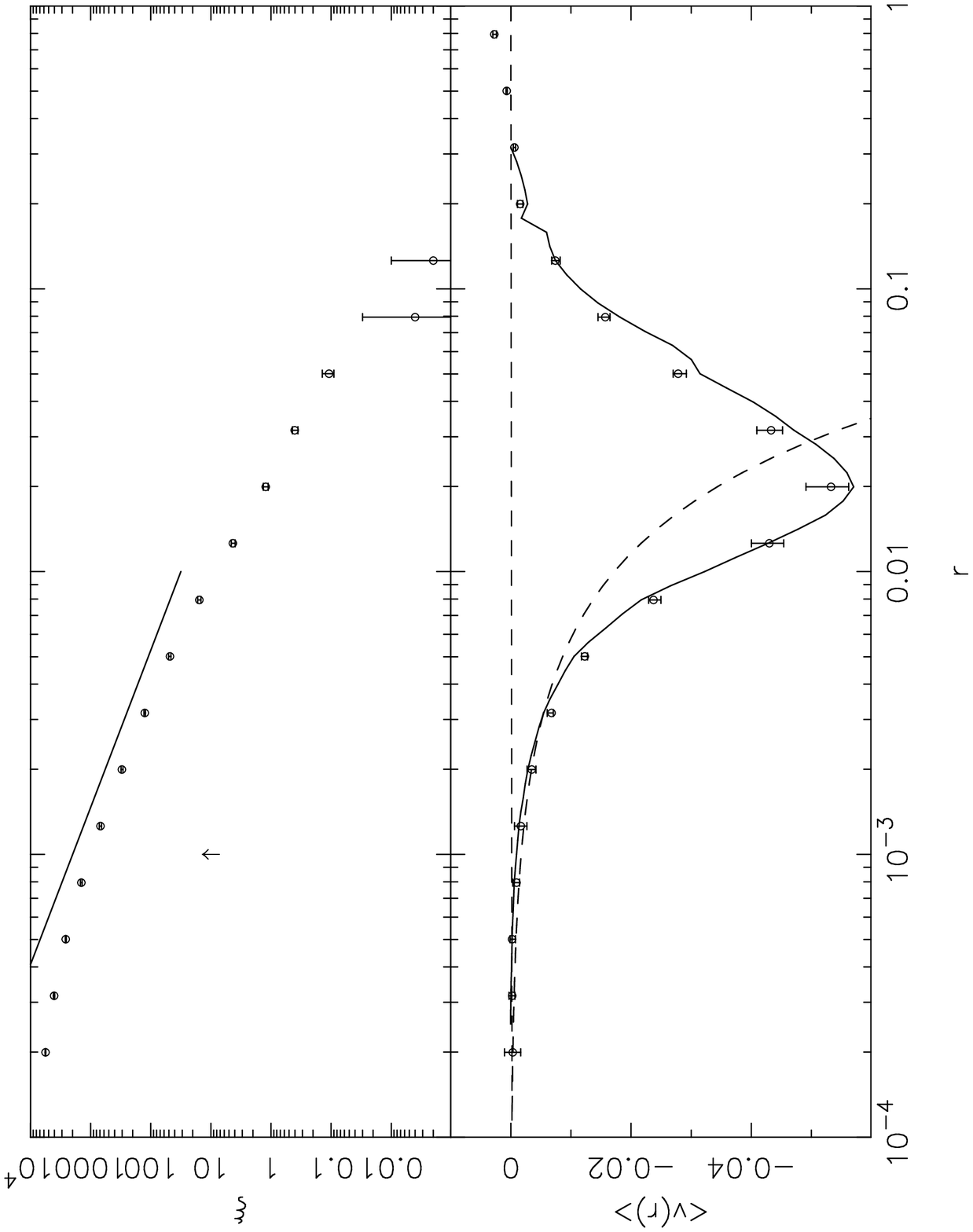]{Scaling test for the
renormalization solution for $n=0$, as in Fig. \ref{fig3}.\label{fig4}}

\figcaption[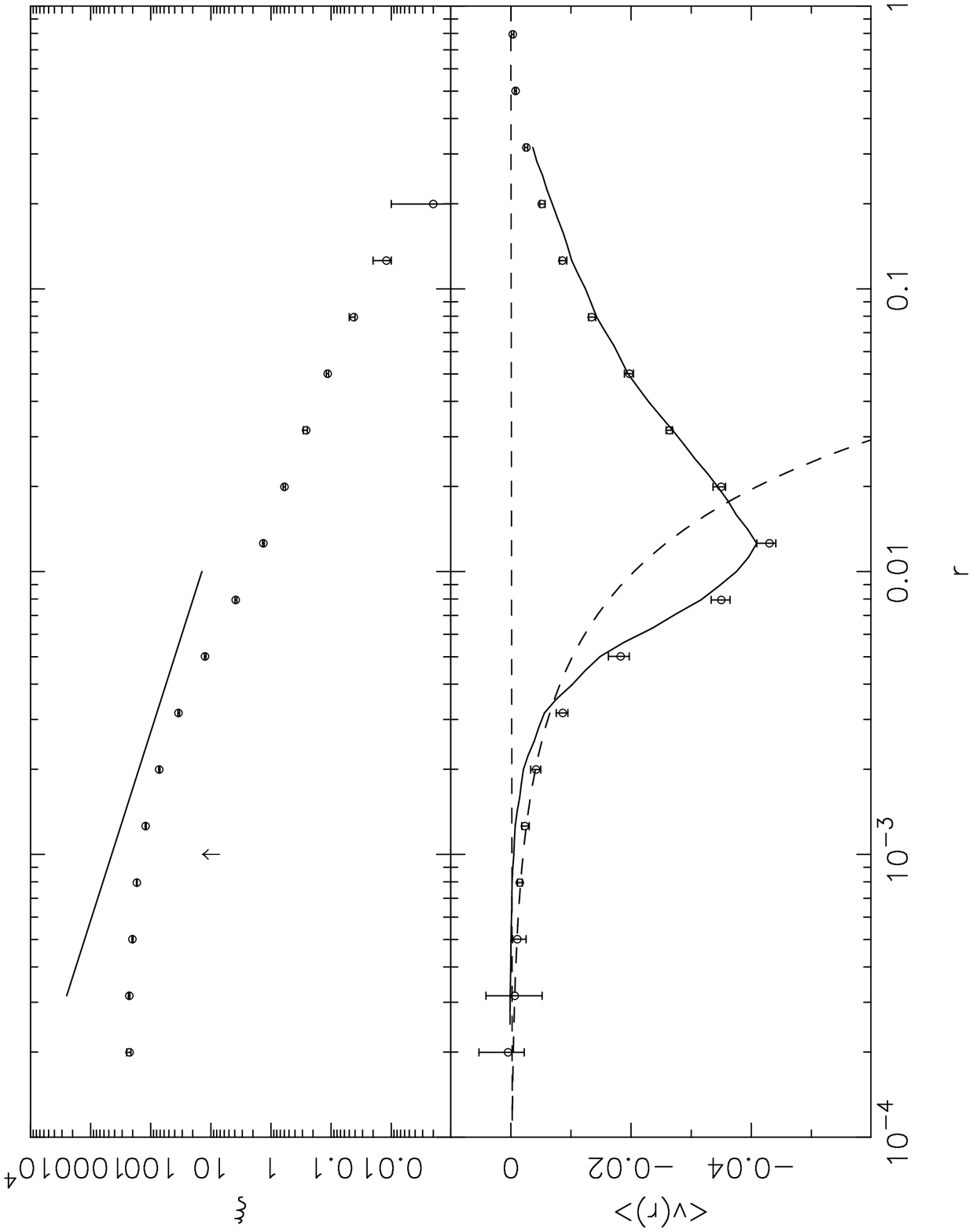]{Scaling test for the conventional
solution for $n=-1$, as in Fig. \ref{fig3}.\label{fig5}}

\figcaption[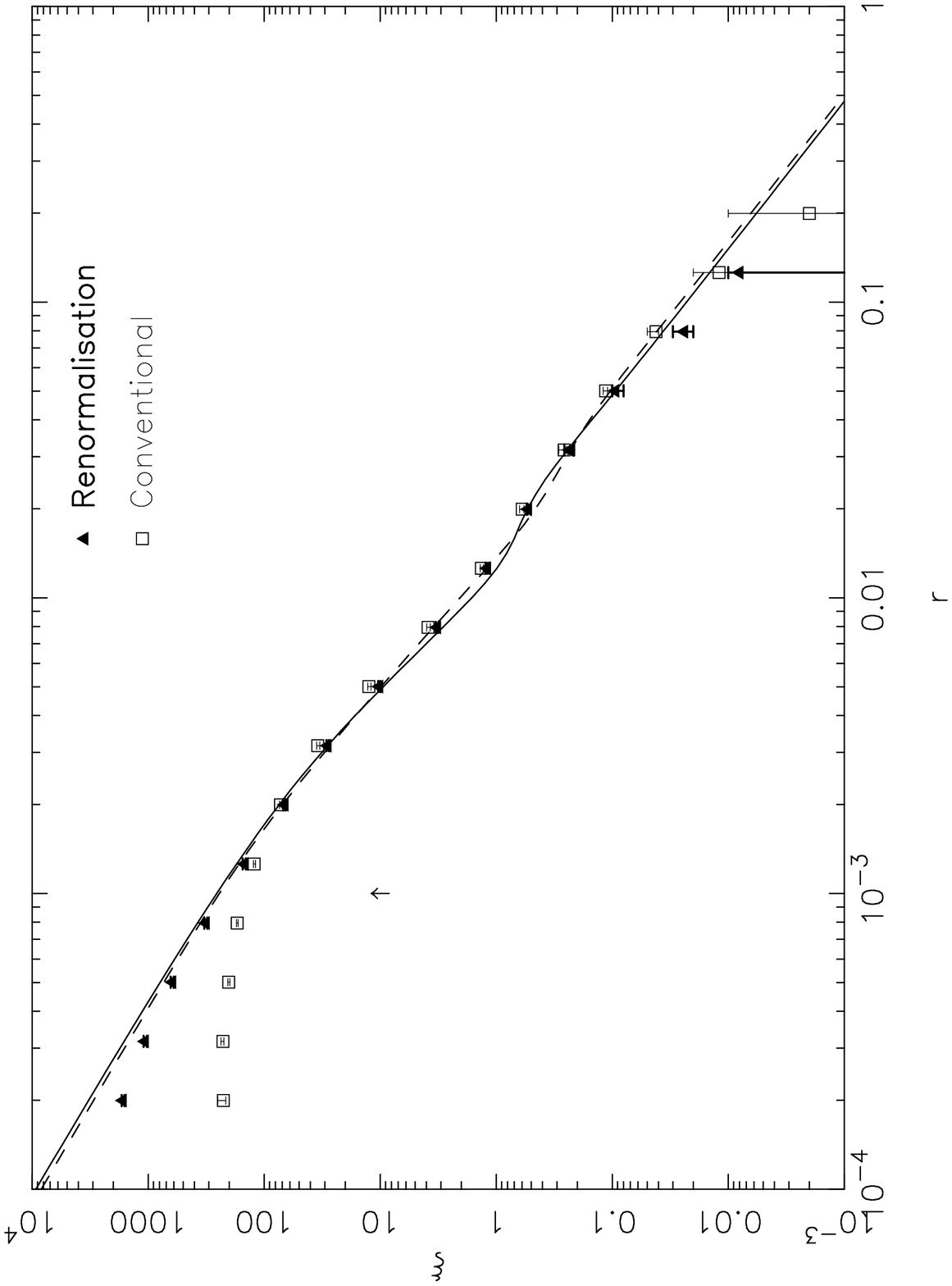]{Comparison of the Hamilton {\it et al.}
(1991) interpolation of the two-point function $\xi (r)$ and the
renormalization and conventional solutions at $n=-1$. The dashed
curve is the fitting function from Peacock \&\ Dodds (1996), and
the solid curve is that of Jain {\it et al.} (1996). The arrow
marks the force law cutoff length $c$.\label{fig6}} 

\figcaption[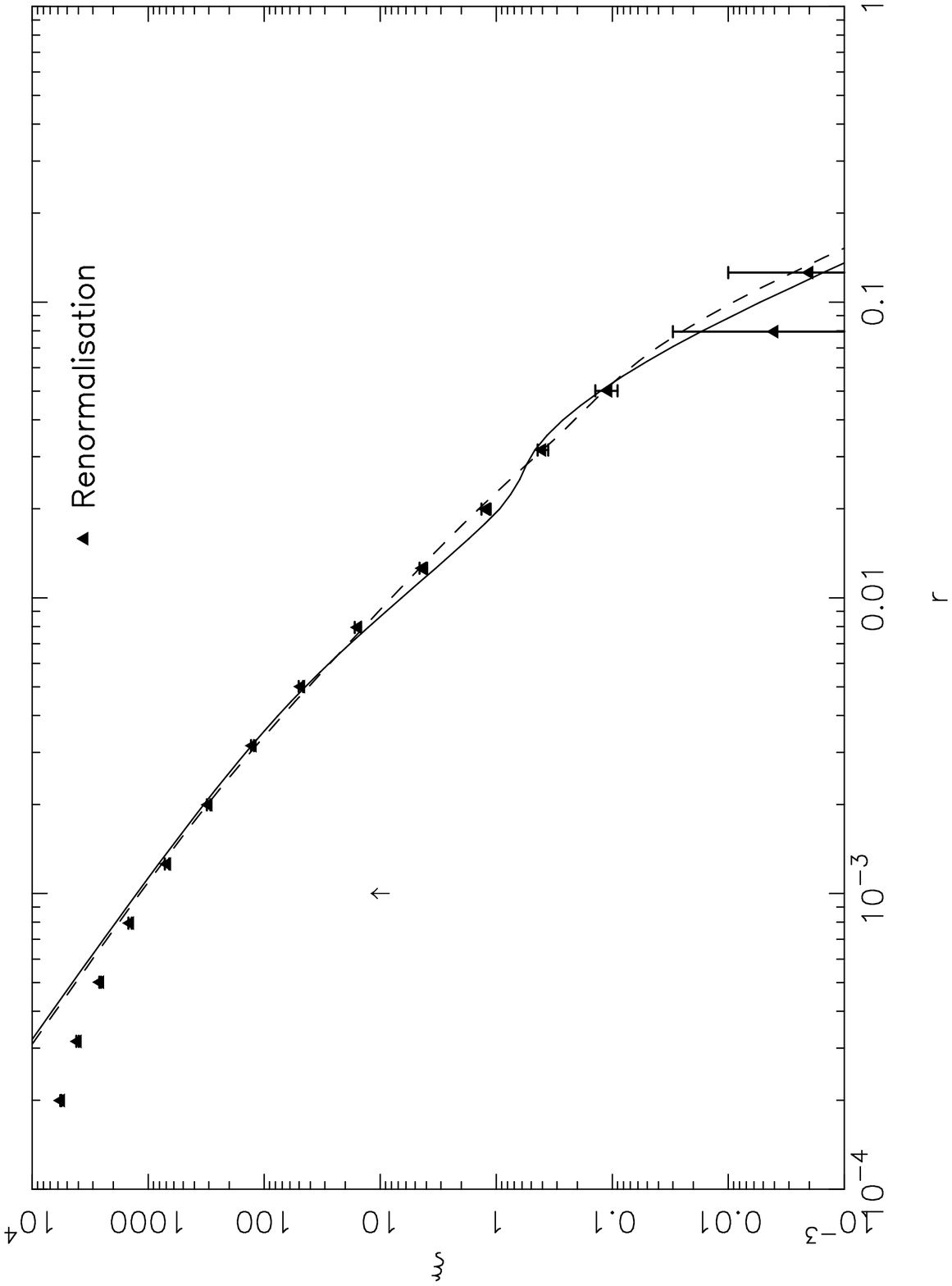]{Comparison of the two-point correlation
function in the renormalization solution for $n=0$ and the
Hamilton {\it et al.} interpolation formula, as in
Fig.~\ref{fig6}.\label{fig7}} 

\figcaption[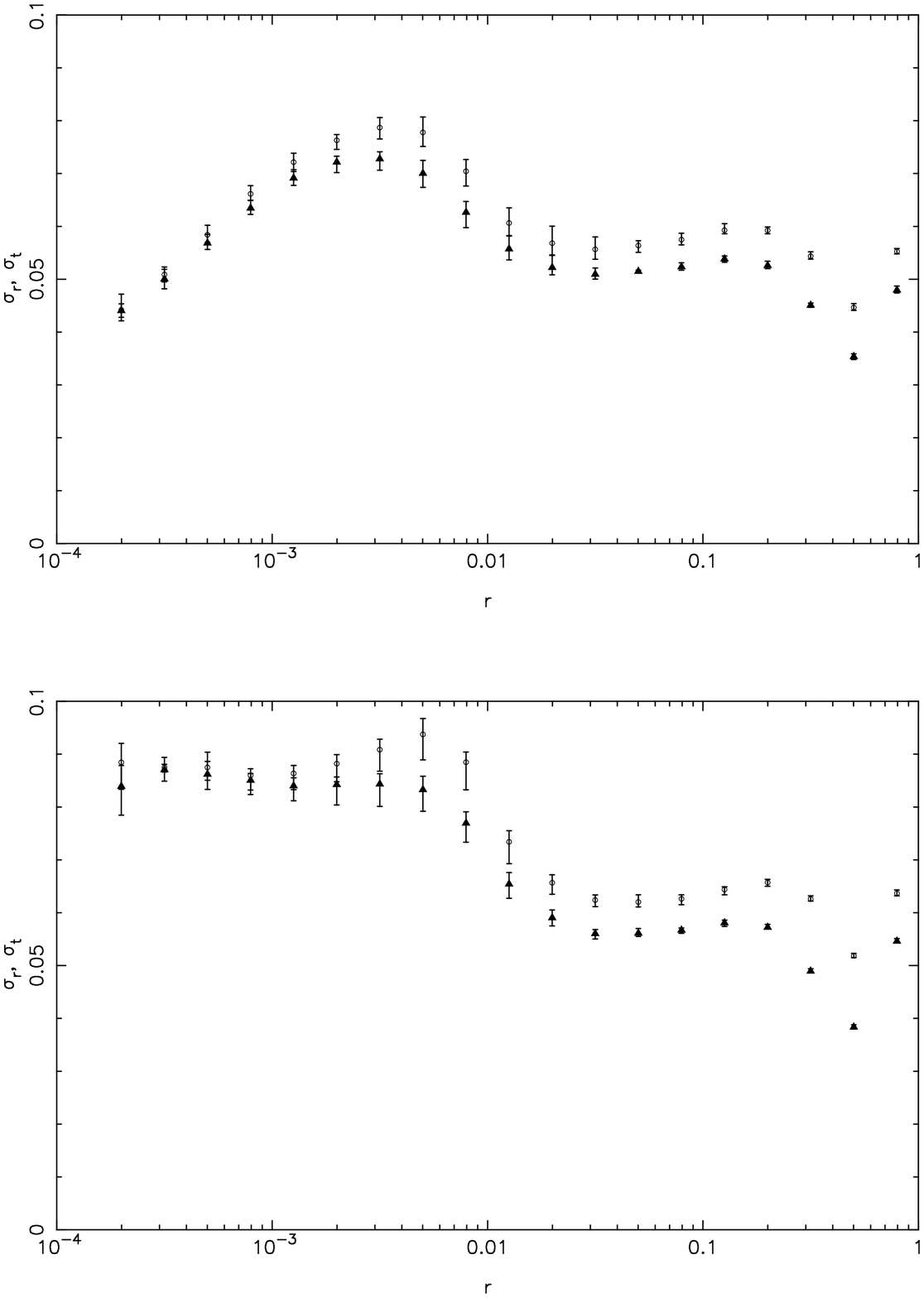]{Relative peculiar velocity dispersions
for $n=-1$. The top panel is the standard renormalization
solution, and the bottom is the comparison conventional
solution. The circles are $\sigma _r$, the rms fluctuation around
the mean of the radial component of the relative peculiar
velocity of particle pairs. The triangles are the rms value
$\sigma _t$ of the relative velocity transverse to the 
line connecting the particles, and normalized to one component,
so in a isotropic distribution $\sigma _t=\sigma _r$. \label{fig8}} 

\figcaption[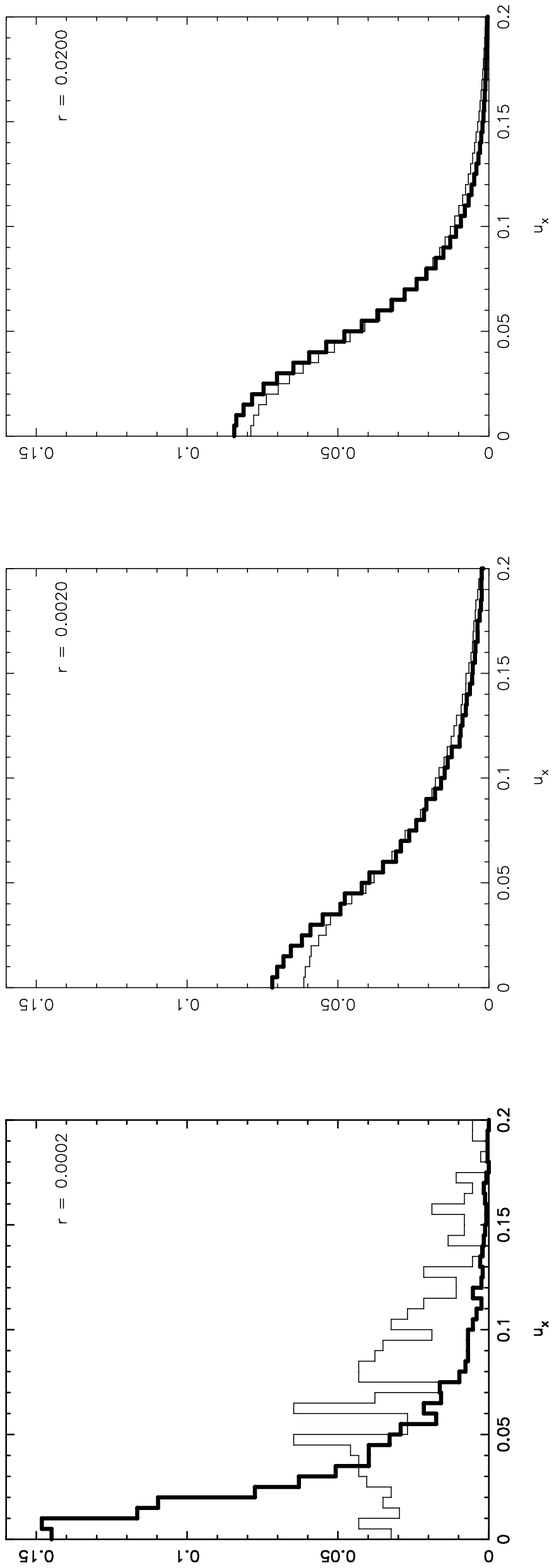]{Frequency distributions in the absolute value
of one Cartesian component of the relative proper velocity difference
of particle pairs, in three bins of separation each of width $\delta
r/r\sim 0.5$. The bold histogram is the standard renormalization
solution, and the thinner histogram is the comparison conventional
solution, for the ``phase-matched'' pair of realizations for
$n=-1$. \label{fig9}} 

\figcaption[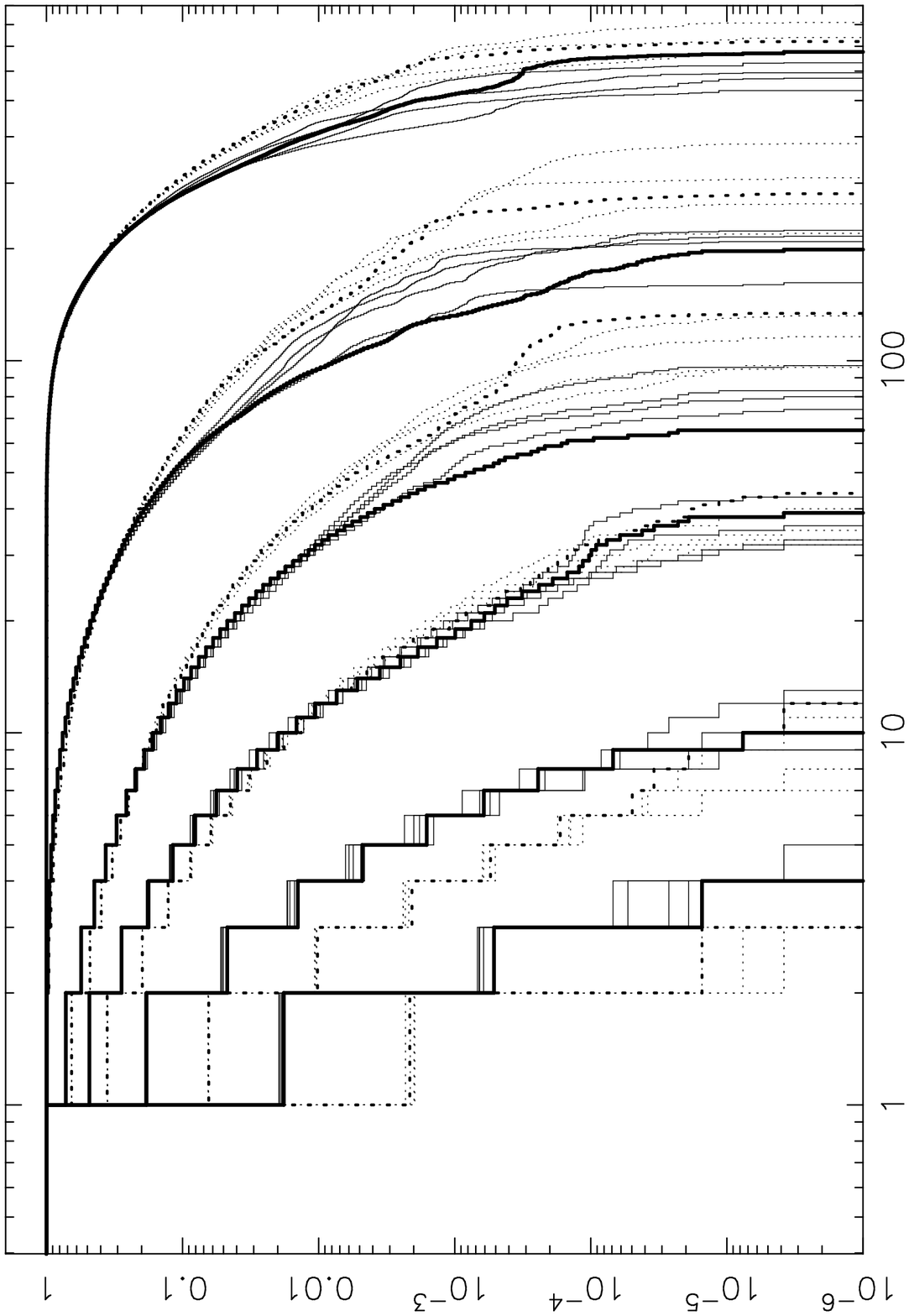]{Cumulative frequency distributions of the
counts of neighboring particles in spheres centered on each of the
$N=64^3$ particles in the $n=-1$ realizations. Histograms are
shown (from left to right) for six 
different sphere radii: $r=$ 0.0002, 0.0007, 0.002, 0.005, 0.02 and
0.05. For each radius the solid histograms are the counts in the
renormalization realizations and the dotted histograms the counts in
the conventional realizations. The ``phase-matched'' pair are
plotted as the heavier solid and dotted lines. 
The cumulative number of neighbors
in the abscissa is plotted at the left-hand edge of each bar in the
histogram.\label{fig10}} 

\figcaption[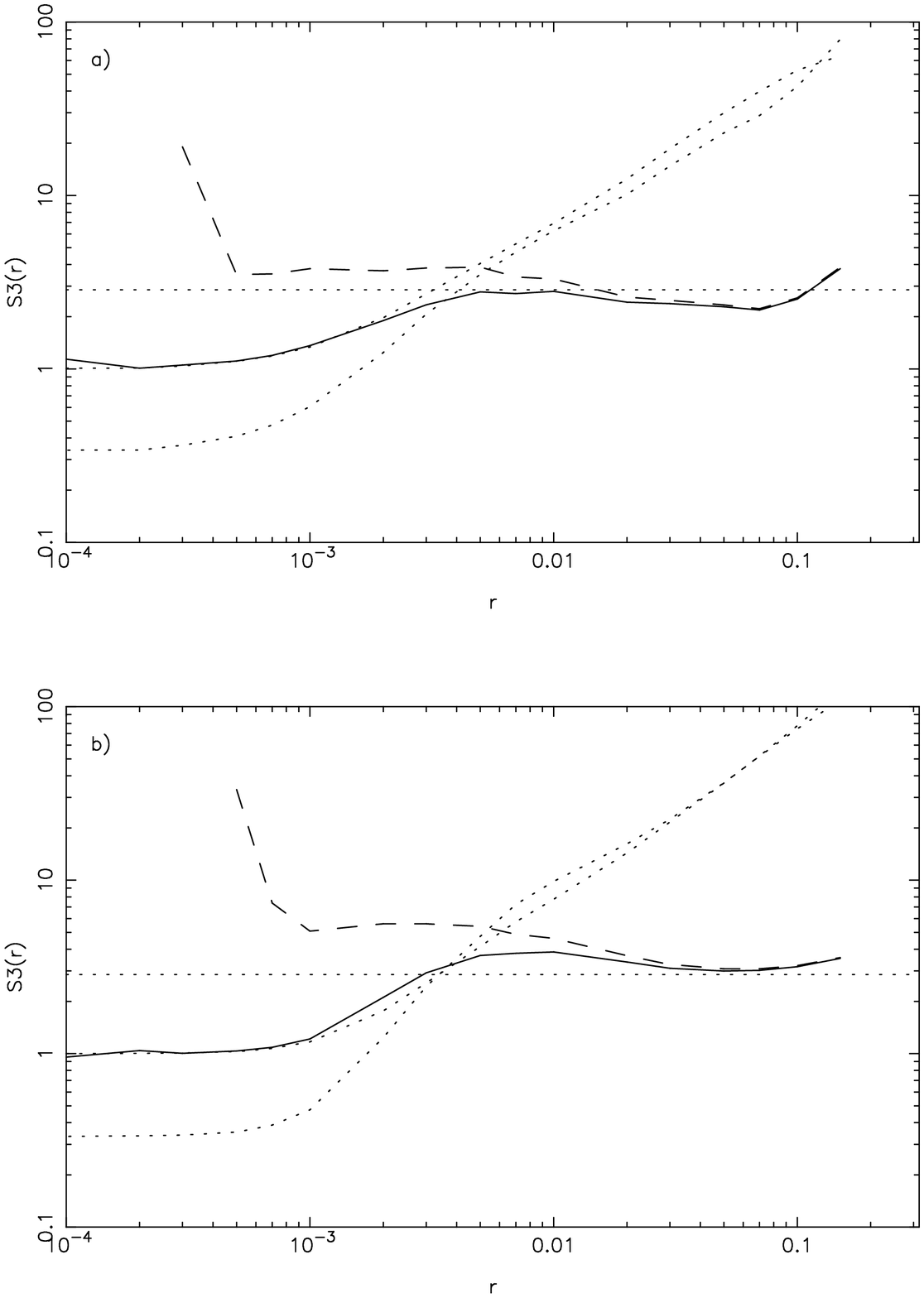]{Measures of the third moments of the
mass distribution, based on the distribution of counts of
particles in randomly placed spheres for the $n=-1$ simulations. The
number of spheres used 
ranged from $10^8$ for $r\le2\times10^{-3}$ to $10^5$ for
$r\ge3\times10^{-2}$. The upper panel
is the standard renormalization solution, and the lower panel is
the comparison 
conventional solution. The solid line is the dimensionless skewness
ratio $S_3$ of the third central moment to the square of the
variance of the mass 
distribution, as defined in equation~(\ref{eq:S3}). The dashed 
line illustrates the effect of removing the shot noise contributions
to the mass moments, as in equation~(\ref{eq:S3f}). At radii
smaller than plotted for the dashed curves the fluctuations are
off the scale of the graph. Another
measure of the shot noise contribution is the set of ratios in
equation~(\ref{eq:shot}). These ratios are plotted as the dotted
curves that 
asymptote at unity for the second moment and $1/3$ for the third
moment. The horizontal line is the perturbation theory
prediction for $S_3$ (eq. [\ref{eq:Bern}]).\label{fig11}}   

\renewcommand{\footnoterule}{} \renewcommand{\thefootnote}{}
\newdimen\footsize \footsize=\hsize \advance\footsize by -30pt
\def\figlab#1{\footnotetext[0]{\hbox to\footsize{\hfill Fig.~\ref{#1}\hfill}}}

\plotone{figure0.ps}
\figlab{fig0}

\plotone{figure1.ps}
\figlab{fig1}

\plotone{figure2.ps}
\figlab{fig2}

\plotone{figure3.ps}
\figlab{fig3}

\plotone{figure4.ps}
\figlab{fig4}

\plotone{figure5.ps}
\figlab{fig5}

\plotone{figure6.ps}
\figlab{fig6}

\plotone{figure7.ps}
\figlab{fig7}

\plotone{figure8.ps}
\figlab{fig8}

\plotone{figure9.ps}
\figlab{fig9}

\plotone{figure10.ps}
\figlab{fig10}

\plotone{figure11.ps}
\figlab{fig11}

\end{document}